# A characterization of continuous variable entanglement

W. P. Bowen, R. Schnabel†, and P. K. Lam
*Department of Physics, Faculty of Science, Australian National University, ACT 0200, Australia*

T. C. Ralph
*Department of Physics, University of Queensland, St Lucia, QLD 4072, Australia*

We present an experimental analysis of quadrature entanglement produced from a pair of amplitude squeezed beams. The correlation matrix of the state is characterized within a set of reasonable assumptions, and the strength of the entanglement is gauged using measures of the degree of inseparability and the degree of EPR paradox. We introduce controlled decoherence in the form of optical loss to the entangled state, and demonstrate qualitative differences in the response of the degrees of inseparability and EPR paradox to this loss. The entanglement is represented on a photon number diagram that provides an intuitive and physically relevant description of the state. We calculate efficacy contours for several quantum information protocols on this diagram, and use them to predict the effectiveness of our entanglement in those protocols.

## I. INTRODUCTION

*Entanglement* is one of the most intriguing features of quantum mechanics. It was first discussed by Einstein, Podolsky, and Rosen in 1935 [1] who used the concept to propose that either quantum mechanics was incomplete or local realism was false. Since that seminal paper experiments have shown entanglement to be a real property of the physical world [2]. Interest in entanglement has grown recently due to its apparent usefulness as an enabling technology in quantum information and communication protocols such as quantum teleportation [3], dense coding [4, 5] and quantum computation [6]. The specific properties of the entangled state utilized in each of these protocols plays a highly significant role in the success of the protocol. It is therefore important to be able to perform complete and accurate characterizations of an available entanglement resource, which is the topic of this paper.

We report the generation and characterization of Gaussian continuous variable entanglement between the amplitude and phase quadratures of a pair of light beams; henceforth termed *quadrature entanglement*. This entanglement has been reported previously [7], the purpose of this paper is to present new experimental results, to more fully characterize the entanglement, and to elaborate on the results presented in that letter. It is well known that Gaussian entanglement can be fully characterized by the coherent amplitudes of the entangled beams, and a matrix containing the correlations between each of the variables of interest (in our case the amplitude and phase quadratures of both entangled beams), termed the *correlation matrix*. To our knowledge, although previously there have been a number of experiments on continuous variable entanglement [8–12], none performed this characterization. Given some reasonable assumptions about our entanglement, we do so here.

Although the coherent amplitudes of the entangled beams and the correlation matrix together provide a complete characterization of quadrature entanglement, they do not directly yield a measure for the strength of the entanglement. In past experiments the strength of an entangled resource has been characterized in the spirit of either the Schrödinger [10–12] or Heisenberg pictures [8, 9, 11], and the characterizations lead to qualitatively different results. In the Schrödinger picture, a necessary and sufficient criterion for the entanglement of a pair of sub-systems is that the state describing the entire system is *inseparable*. That is, it is not possible to factor the wavefunction of the entire system into a product of separate contributions from each sub-system. Given that an observable signature of the mathematical criterion for wave-function entanglement can be identified, one can define the *degree of inseparability* for the state, and use it to characterize the strength of the entanglement. In the Heisenberg picture, a sufficient criterion for entanglement is that correlations between conjugate observables of two sub-systems allow the statistical inference of either observable in one sub-system, upon a measurement in the other, to be smaller than the standard quantum limit. That is, the presence of non-classical correlations. This approach was originally proposed by Einstein, Podolsky and Rosen [1] and has since been termed the *EPR paradox*. Similarly to the Schrödinger picture we can define the *degree of EPR paradox* for a given entangled state, and use it to characterize the strength of the entanglement. For pure states the Schrödinger and Heisenberg approaches return qualitatively equivalent results suggesting consistency of the two methods. However, when decoherence is present, causing the state to be mixed, differences can occur. For quadrature entanglement wave-function inseparability may be identified using the *inseparability criterion* proposed by Duan *et al.* [13, 14]. We use this criterion to define the degree of inseparability of our entanglement. To define the degree of EPR paradox we use the criterion for demonstration of the EPR paradox as quantified by Reid and Drummond [15], and refer to this as the *EPR paradox criterion*. By introducing decoherence in the form of optical loss to both of our entangled beams we observe qualitative differences between the degree of inseparability and the degree of EPR paradox.

---

† now at the Institut für Atom- und Molekülphysik, Universität Hannover, 30167 Hannover, Germany.

Finally, we characterize our entanglement in terms of mean sideband photon numbers [7]. We find that the mean number of photons per bandwidth per time in the sidebands of an entangled state can be broken into four categories: the mean number of photons required to maintain the entanglement, to produce any bias that exists between the amplitude and phase quadratures of the beams, to produce the impurity of the state, and to produce any impurity bias between the amplitude and phase quadratures. For our entanglement, these four mean photon numbers provide an equivalent but more intuitive characterization to the correlation matrix. We attach less significance to the mean photon numbers resulting from impurity than those required to maintain and bias the entanglement, and sum them to give the total mean photon number per bandwidth per time due to impurity. Our entanglement could then be represented on a three dimensional photon number diagram. On a plane of this diagram, we directly assessed the level of success achievable for quantum teleportation, demonstration of the EPR paradox, and high and low photon number dense coding when utilizing our entanglement. The photon number diagram can also be used to assess the effect of techniques such as distillation and purification, that can be used to improve the quality of an entangled state.

## II. PRODUCTION OF CONTINUOUS VARIABLE ENTANGLEMENT

In the time domain, a single mode of the electro-magnetic field can be fully defined by its field annihilation operator $\tilde{a}(t)$, which has the commutation relation $[\tilde{a}(t), \tilde{a}^\dagger(t)] = 1$. $\tilde{a}(t)$ is non-Hermitian but can be expanded as

$$\tilde{a}(t) = \alpha(t) + \frac{\delta\tilde{X}^+(t) + i\delta\tilde{X}^-(t)}{2}, \qquad (1)$$

where $\delta\tilde{X}^\pm(t)$ are the time domain Hermitian amplitude (super-script +) and phase (super-script -) quadrature noise operators, and $\alpha(t) = \langle\tilde{a}(t)\rangle$ is the coherent amplitude of the field which we define to be real throughout this paper without loss of generality. The commutation relation $[\tilde{X}^+(t), \tilde{X}^-(t)] = 2i$ follows directly from the commutation relation of $\tilde{a}(t)$ and $\tilde{a}^\dagger(t)$. This relation places a fundamental limitation on how well one quadrature of an optical beam can be known, given some knowledge of the orthogonal quadrature. This can be expressed as the uncertainty product $\Delta^2\tilde{X}^+(t)\Delta^2\tilde{X}^-(t) > 1$, where the operator variances are denoted by $\Delta^2\tilde{X} = \langle(\delta\tilde{X})^2\rangle$. It is this uncertainty product that makes quadrature entanglement possible.

Several techniques may be used to generate quadrature entanglement. It was first generated by Ou *et al.* in 1992 [8, 9] using a non-degenerate optical parametric amplifier, and more recently using the Kerr non-linearity in fibers [11], and interfering the outputs of two below threshold optical parametric amplifiers [7, 12, 16]. Ultimately all of these techniques yield Gaussian continuous variable entanglement of a form that can be modelled simply and, as we will see in section III, quite generally, by combining two quadrature squeezed beams with orthogonal squeezing on a 50/50 beam splitter. Indeed, it is this technique that we adopted to experimentally generate quadrature entanglement. In general, the two beam splitter outputs $\tilde{a}_x(t)$ and $\tilde{a}_y(t)$ are of the form

$$\tilde{a}_x(t) = \frac{e^{i\phi_x}}{\sqrt{2}}\left(\tilde{a}_{\rm sqz,1}(t) + e^{i\theta}\tilde{a}_{\rm sqz,2}(t)\right) \qquad (2)$$

$$\tilde{a}_y(t) = \frac{e^{i\phi_y}}{\sqrt{2}}\left(\tilde{a}_{\rm sqz,1}(t) - e^{i\theta}\tilde{a}_{\rm sqz,2}(t)\right), \qquad (3)$$

where $\tilde{a}_{\rm sqz,1}(t)$ and $\tilde{a}_{\rm sqz,2}(t)$ are the annihilation operators of the input squeezed beams, $\theta$ defines the relative phase between them, $\phi_x$ and $\phi_y$ are phase shifts that rotate the operators such that $\alpha_x(t)$ and $\alpha_y(t)$ are real, and throughout this paper the sub-scripts $x$ and $y$ denote the beams being interrogated for entanglement. To avoid frequency dependant noise sources present on our optical fields we examine our entangled states in the frequency domain. The transfer from time to frequency domain can be achieved simply by taking a Fourier transform. Henceforth, we perform this transform and distinguish operators in the frequency domain by replacing the decoration $\tilde{}$ with a $\hat{}$. For conciseness where possible we omit the frequency domain functionality $(\omega)$. We have already taken the time domain coherent amplitude of the our optical fields to be real, but this property does not carry over to the frequency domain. We denote the real and imaginary parts of the frequency domain coherent amplitude respectively as $\alpha^+ = \Re\{\alpha(\omega)\} = 2\langle\hat{X}^+\rangle$ and $\alpha^- = \Im\{\alpha(\omega)\} = 2\langle\hat{X}^-\rangle$. We take the input beams to be amplitude squeezed states ($\Delta^2\hat{X}^+_{\rm sqz,1} < 1$ and $\Delta^2\hat{X}^+_{\rm sqz,2} < 1$) with equal intensities ($\alpha_{\rm sqz,1}(t) = \alpha_{\rm sqz,2}(t)$), and set $\theta = \pi/2$ so that the squeezed quadratures are orthogonal at the beam splitter. The frequency domain amplitude and phase quadratures of the output beams $x$ and $y$ can then be expressed as

$$\hat{X}^\pm_x = \frac{1}{2}\left(\pm\hat{X}^+_{\rm sqz,1} + \hat{X}^+_{\rm sqz,2} + \hat{X}^-_{\rm sqz,1} \mp \hat{X}^-_{\rm sqz,2}\right) \qquad (4)$$

$$\hat{X}^\pm_y = \frac{1}{2}\left(\hat{X}^+_{\rm sqz,1} \pm \hat{X}^+_{\rm sqz,2} \mp \hat{X}^-_{\rm sqz,1} + \hat{X}^-_{\rm sqz,2}\right). \qquad (5)$$

We see that as the squeezing of the input beams approaches perfect ($\{\Delta^2\hat{X}^+_{\rm sqz,1}, \Delta^2\hat{X}^+_{\rm sqz,2}\} \to 0$) the quadrature noise operators of beams $x$ and $y$ approach

$$\delta\hat{X}^\pm_x \to \frac{1}{2}\left(\delta\hat{X}^-_{\rm sqz,1} \mp \delta\hat{X}^-_{\rm sqz,2}\right) \qquad (6)$$

$$\delta\hat{X}^\pm_y \to \mp\frac{1}{2}\left(\delta\hat{X}^-_{\rm sqz,1} \mp \delta\hat{X}^-_{\rm sqz,2}\right), \qquad (7)$$

so that

$$\left\langle \left(\delta\hat{X}_x^+ + \delta\hat{X}_y^+\right)^2 \right\rangle \to 0 \quad (8)$$

$$\left\langle \left(\delta\hat{X}_x^- - \delta\hat{X}_y^-\right)^2 \right\rangle \to 0. \quad (9)$$

Therefore in this limit an amplitude quadrature measurement on beam $x$ would provide an exact prediction of the amplitude quadrature of beam $y$; and similarly a phase quadrature measurement on beam $x$ would provide an exact prediction of the phase quadrature of beam $y$. This is a demonstration of the EPR paradox in exactly the manner proposed in the seminal paper of Einstein *et al.* [1]. Analysis of the entanglement in the physically realistic regime where $\{\Delta^2\hat{X}^+_{\rm sqz,1}, \Delta^2\hat{X}^+_{\rm sqz,2}\} \neq 0$ is more complex, and is the topic of the following section.

## III. CHARACTERIZATION OF CONTINUOUS VARIABLE ENTANGLEMENT

Characterization of continuous variable entanglement is, in many ways, a more complex enterprise than its discrete variable counterpart. Discrete variable entanglement can be fully characterized by a density matrix of finite dimension (usually 4×4). In contrast, complete characterization of continuous variable entanglement requires a density matrix of infinite size. This problem has received considerable interest in the quantum optics community with, as of now, no consensus on the most appropriate characterization method [17]. However, experimental realizations of continuous variable entanglement have, to date, been limited to a sub-class of states - those with Gaussian statistics - for which well defined characterization techniques do exist. In this section we introduce the characterization techniques used for our entanglement, and discuss a new interpretation separating the mean number of photons per bandwidth per time in the entangled state into components required to maintain and bias the entangled state, and to produce and bias the impurity present in the state [7, 18].

### A Gaussian entanglement and the correlation matrix

Any Gaussian continuous variable bi-partite state can be fully characterized by its amplitude and phase quadrature coherent amplitudes $\alpha_x^\pm$, $\alpha_y^\pm$, and the correlation (or covariance) matrix. In general $\alpha_x^\pm$ and $\alpha_y^\pm$ are easily characterized, and do not contribute to the strength of entanglement exhibited by the state. In our experiment the entangled state was produced from two squeezed vacuum states, so that the amplitude and phase quadrature coherent amplitudes of beams $x$ and $y$ were all zero, $\alpha_x^\pm = \alpha_y^\pm = 0$. We will therefore focus on the correlation matrix here. The correlation matrix $CM$ is given by

$$CM = \begin{pmatrix} C_{xx}^{++} & C_{xx}^{+-} & C_{xy}^{++} & C_{xy}^{+-} \\ C_{xx}^{-+} & C_{xx}^{--} & C_{xy}^{-+} & C_{xy}^{--} \\ C_{yx}^{++} & C_{yx}^{+-} & C_{yy}^{++} & C_{yy}^{+-} \\ C_{yx}^{-+} & C_{yx}^{--} & C_{yy}^{-+} & C_{yy}^{--} \end{pmatrix}. \quad (10)$$

Each term in this matrix is the correlation co-efficient between two of the variables $\hat{X}_x^+$, $\hat{X}_x^-$, $\hat{X}_y^+$, and $\hat{X}_y^-$; defined as

$$C_{mn}^{kl} = \frac{1}{2}\left\langle \hat{X}_m^k\hat{X}_n^l + X_n^l\hat{X}_m^k \right\rangle - \left\langle \hat{X}_m^k \right\rangle \left\langle \hat{X}_n^l \right\rangle \quad (11)$$

$$= \frac{1}{2}\left\langle \delta\hat{X}_m^k\delta\hat{X}_n^l + \delta\hat{X}_n^l\delta\hat{X}_m^k \right\rangle, \quad (12)$$

with $\{k,l\} \in \{+,-\}$, $\{m,n\} \in \{x,y\}$. The symmetry in the form of $C_{mn}^{kl}$ dictates that in general $C_{mn}^{kl} = C_{nm}^{lk}$. The correlation matrix is therefore fully specified by ten independent co-efficients.

The entangled beams analyzed in this paper were generated in a symmetric manner by interfering two amplitude squeezed beams with $\pi/2$ phase shift on a 50/50 beam splitter (as discussed in the previous section), and encountered identical loss before detection. Furthermore, the squeezed beams themselves were produced in an identical manner in identical OPAs, with no cross quadrature correlations present either within each beam individually or between the beams. When applied to eqs. (4) and (5) these symmetries dictate that the amplitude (phase) quadrature variances of beams $x$ and $y$ are equal, $\Delta^2\hat{X}^\pm = \Delta^2\hat{X}_x^\pm = \Delta^2\hat{X}_y^\pm$, so that $C_{mm}^{\pm\pm} = \Delta^2\hat{X}^\pm$; and that the beams exhibit no cross-quadrature correlations. That is, that $C_{mn}^{\pm\mp} = 0$. The correlation matrix is then given by

$$CM = \begin{pmatrix} C_{xx}^{++} & 0 & C_{xy}^{++} & 0 \\ 0 & C_{xx}^{--} & 0 & C_{xy}^{--} \\ C_{xy}^{++} & 0 & C_{xx}^{++} & 0 \\ 0 & C_{xy}^{--} & 0 & C_{xx}^{--} \end{pmatrix}, \quad (13)$$

where complete specification now only requires characterization of $\Delta^2\hat{X}^+$, $\Delta^2\hat{X}^-$, $\langle\delta\hat{X}_x^+\delta\hat{X}_y^+ + \delta\hat{X}_y^+\delta\hat{X}_x^+\rangle$, and $\langle\delta\hat{X}_x^-\delta\hat{X}_y^- + \delta\hat{X}_y^-\delta\hat{X}_x^-\rangle$. Specification of these four parameters is equivalent to characterization of the variance of the squeezed and anti-squeezed quadratures of the pair of squeezed beams produced by re-combining the entangled beams losslessly and in-phase on a 50/50 beam splitter.

### B The inseparability criterion

Specification of the correlation matrix, although it does offer a complete description of the entanglement, does not immediately provide a measure of whether beams $x$ and $y$ are entangled, or how strongly they are entangled. We use two criteria, both of which can be inferred from the correlation matrix, to measure those properties. In this section we discuss the Inseparability criterion recently proposed by Duan *et al.* [13, 14] which provides a necessary and sufficient condition for Gaussian entanglement; and in the section following we introduce the EPR paradox criterion proposed by Reid and Drummond [15] which has been used to characterize entanglement in past experiments. It should be noted that strictly speaking, a good measure of entanglement should satisfy the conditions given in [19, 20], and stated explicitly later in this paper. Neither the inseparability or EPR criteria have been shown to satisfy these conditions, and indeed, to our knowledge no such measure exists presently for continuous variable entanglement. However, both criteria considered here have strong physical significance, have a straight forward dependance on the strength of the quantum resources used to generate the entanglement, and are commonly used to gauge the strength of entanglement in experiments. Throughout this paper we, therefore, refer to both criteria as measures of the strength of entanglement.

The inseparability criterion relies on the identification of separability with positivity of the P-representation distribution of the state. Duan *et. al*[13] showed that through local linear unitary Bogoliubov operations any bi-partite Gaussian state can be transformed so that its correlation matrix has the standard form

$$CM_s = \begin{pmatrix} C_{xx}^{++} & 0 & C_{xy}^{++} & 0 \\ 0 & C_{xx}^{--} & 0 & C_{xy}^{--} \\ C_{xy}^{++} & 0 & C_{yy}^{++} & 0 \\ 0 & C_{xy}^{--} & 0 & C_{yy}^{--} \end{pmatrix}, \tag{14}$$

where the values of $C_{nm}^{\pm\pm}$ are restricted by the conditions

$$\frac{C_{xx}^{++}-1}{C_{yy}^{++}-1} = \frac{C_{xx}^{--}-1}{C_{yy}^{--}-1}, \tag{15}$$

and

$$\sqrt{(C_{xx}^{++}-1)(C_{yy}^{++}-1)} - |C_{xy}^{++}| = \sqrt{(C_{xx}^{--}-1)(C_{yy}^{--}-1)} - |C_{xy}^{--}|. \tag{16}$$

Given that the state is in this form, they showed that the Inseparability criterion

$$\Delta^2\hat{X}_I^+ + \Delta^2\hat{X}_I^- < 2\left(k^2 + \frac{1}{k^2}\right), \tag{17}$$

is a necessary and sufficient condition for the presence of entanglement[13], where $\Delta^2\hat{X}_I^\pm$ are the measurable correlations

$$\Delta^2\hat{X}_I^\pm = \left\langle \left( k\delta\hat{X}_x^\pm - \frac{C_{xy}^{\pm\pm}}{|C_{xy}^{\pm\pm}|}\frac{\delta\hat{X}_y^\pm}{k}\right)^2 \right\rangle, \tag{18}$$

and $k$ is a parameter that compensates for bias between subsystems $x$ and $y$ and is given by

$$k = \left(\frac{C_{yy}^{++}-1}{C_{xx}^{++}-1}\right)^{\frac{1}{4}} = \left(\frac{C_{yy}^{--}-1}{C_{xx}^{--}-1}\right)^{\frac{1}{4}}. \tag{19}$$

In fact, Duan *et. al* showed that if the state under interrogation is separable satisfaction of criterion (17) is impossible for any arbitrary $k$. From an experimental perspective $k$ can then be thought of as a variable parameter. Satisfaction of the criterion for any $k$ is a sufficient condition for entanglement.

A comparison of the form of the correlation matrix describing our entanglement (eq. (13)) with Duan *et al.*'s standard form (eqs. (14), (15), and (16)) reveals that, in general, we cannot directly apply the Inseparability criterion of eq. (17). Of course, after a complete characterisation of the correlation matrix it can be taken into the standard form, and the Inseparability criterion can then be applied. However, we will see in the following analysis that if a product form of the criterion is taken, it becomes valid for a wider range of correlation matrices and indeed is then directly applicable to our entanglement. Let us consider the effect that restrictions (15) and (16) have on $\Delta^2\hat{X}_I^\pm$. Expanding $\Delta^2\hat{X}_I^\pm$ we find

$$\Delta^2\hat{X}_I^\pm = k^2\Delta^2 X_x^\pm + \frac{\Delta^2 X_y^\pm}{k^2} - 2\frac{C_{xy}^{\pm\pm}}{|C_{xy}^{\pm\pm}|}\left\langle \delta X_x^\pm \delta X_y^\pm \right\rangle \tag{20}$$

$$= \sqrt{\frac{C_{yy}^{\pm\pm}-1}{C_{xx}^{\pm\pm}-1}}C_{xx}^{\pm\pm} + \sqrt{\frac{C_{xx}^{\pm\pm}-1}{C_{yy}^{\pm\pm}-1}}C_{yy}^{\pm\pm} - 2\left|C_{xy}^{\pm\pm}\right|$$

$$= 2\left[\sqrt{\left(C_{xx}^{\pm\pm}-1\right)\left(C_{yy}^{\pm\pm}-1\right)} - \left|C_{xy}^{\pm\pm}\right|\right] + \sqrt{\frac{C_{xx}^{\pm\pm}-1}{C_{yy}^{\pm\pm}-1}} + \sqrt{\frac{C_{yy}^{\pm\pm}-1}{C_{xx}^{\pm\pm}-1}}. \tag{21}$$

A comparison of eqs. (21) with restrictions (15) and (16) reveals that transforming a general bi-partite Gaussian state into the standard form for which the inseparability criteria of eq. (17) is valid equates $\Delta^2\hat{X}_I^+$ and $\Delta^2\hat{X}_I^-$ ($\Delta^2\hat{X}_I^+ = \Delta^2\hat{X}_I^-$). The inseparability criteria can therefore be equivalently written in the product form

$$\sqrt{\Delta^2\hat{X}_I^+\Delta^2\hat{X}_I^-} < \left(k^2 + \frac{1}{k^2}\right). \tag{22}$$

In this form however, the criterion is insensitive to equal local squeezing operations on beams $x$ and $y$. This was not the case for the sum criterion, where it was necessary that restrictions (15) and (16) forbid those operations. The product form of the inseparability criterion is therefore valid for a wider set of correlations matrices. Indeed we find that validity of the product form only requires one restriction on the form of the correlation matrix, rather than the two in eqs. (15) and (16). This restriction can be shown to be

$$C_{yy}^{++}C_{xx}^{--} - C_{xx}^{++}C_{yy}^{--} = \sqrt{\frac{\Delta^2\hat{X}_I^-}{\Delta^2\hat{X}_I^+}}\left(C_{yy}^{++} - C_{xx}^{++}\right) + \sqrt{\frac{\Delta^2\hat{X}_I^+}{\Delta^2\hat{X}_I^-}}\left(C_{xx}^{--} - C_{yy}^{--}\right). \tag{23}$$

Since for our entanglement $C_{xx}^{++} = C_{yy}^{++}$ and $C_{xx}^{--} = C_{yy}^{--}$ (see eq. (13)), we see that this less stringent restriction is satisfied. The correlation matrix describing our entanglement given in eq. (13) is of the same form as that in eq. (14), therefore the product form of the inseparability criterion is directly valid for our entanglement. To provide a direct measure of the strength of the entanglement we define the *degree of inseparability*

$$\mathcal{I} = \frac{\sqrt{\Delta^2\hat{X}_I^+\Delta^2\hat{X}_I^-}}{k^2 + 1/k^2}, \tag{24}$$

normalized such that beams $x$ and $y$ are entangled if $\mathcal{I} < 1$.

For entanglement produced as described in section II the expression for $\mathcal{I}$ becomes considerably simpler. The entangled beams are produced on a 50/50 beam splitter, furthermore, prior to detection they encounter only linear optics and incur equal loss. There is, therefore, symmetry between the quadratures of beams $x$ and $y$, so that $C^{\pm\pm}_{xx} = C^{\pm\pm}_{yy}$. In this case we see from eq. (19) that $k = 1$. Eq. (24) can then be written

$$\mathcal{I} = \sqrt{\Delta^2 \hat{X}^+_{x\pm y} \Delta^2 \hat{X}^-_{x\pm y}}, \tag{25}$$

where $\Delta^2 \hat{O}_{x\pm y}$ is the minimum of the variance of the sum or difference of the operator $\hat{O}$ between beams $x$ and $y$ normalized to the two beam shotnoise, $\Delta^2 \hat{O}_{x\pm y} = \min \langle (\delta\hat{O}_x \pm \delta\hat{O}_y)^2 \rangle / 2$. This measure of entanglement in terms of the product of sum and difference variances between the beams has been used previously in the literature [21].

We are interested in the effect of decoherence in the form of optical loss on the EPR paradox and inseparability criteria, and the photon number diagram. It can be shown from eqs. (4), (5) and (25) that for entanglement generated from a pair of uncorrelated squeezed beams as detailed in section II, and with equal optical loss for beams $x$ and $y$, $\mathcal{I}$ can be expressed as a function of the overall detection efficiency $\eta$ as

$$\mathcal{I} = \eta \Delta^2 \hat{X}^+_{\rm sqz,ave} + (1-\eta), \tag{26}$$

where we define the average of the input beam squeezing as $\Delta^2 \hat{X}^+_{\rm sqz,ave} = (\Delta^2 \hat{X}^+_{\rm sqz,1} + \Delta^2 \hat{X}^+_{\rm sqz,2})/2$. We see that so long as the average squeezing of the two beams used to generate the entanglement is below one ($\Delta^2 \hat{X}^+_{\rm sqz,ave} < 1$), then $\mathcal{I} < 1$. So beams $x$ and $y$ are entangled for any level of input squeezing. Notice that even as $\eta$ approaches zero, for any level of squeezing $\mathcal{I}$ remains below unity. We see that the entanglement is robust against losses at least in the sense that loss alone cannot transform an inseparable state to a separable one.

### C The EPR paradox criterion

The concept of entanglement was first introduced by Einstein, Podolsky, and Rosen in 1935 [1]. They demonstrated than an apparent violation of the Heisenberg uncertainty principle could be achieved between the position and momentum observables of a pair of particles [22]. This apparent violation has since been termed the *EPR paradox*. Demonstration of the EPR paradox relies on quantum correlations between a pair of non-commuting observables, so that measurement of either observable in sub-system $x$ allows the inference of that variable in sub-system $y$ to better than the standard quantum limit. Between the amplitude and phase quadratures of a pair of optical beams this is quantified by the product of conditional variances [15], we therefore define the *degree of EPR paradox* $\mathcal{E}$

$$\mathcal{E} = \Delta^2 \hat{X}^+_{x|y} \Delta^2 \hat{X}^-_{x|y}, \tag{27}$$

where the EPR paradox is demonstrated for $\mathcal{E} < 1$ and the quadrature conditional variances $\Delta^2 \hat{X}^\pm_{x|y}$ are given by

$$\Delta^2 \hat{X}^\pm_{x|y} = \Delta^2 \hat{X}^\pm_x - \frac{\left|\langle \delta X^\pm_x \delta X^\pm_y \rangle\right|^2}{\Delta^2 \hat{X}^\pm_y} \tag{28}$$

$$= C^{\pm\pm}_{xx} - \frac{\left|C^{\pm\pm}_{xy}\right|^2}{C^{\pm\pm}_{xx}} \tag{29}$$

$$= \min_{g^\pm} \left\langle \left(\delta X^\pm_x - g^\pm \delta X^\pm_y\right)^2 \right\rangle, \tag{30}$$

where $g^\pm$ are experimentally adjustable variables. Satisfaction of the EPR paradox criterion between two beams is a sufficient but not necessary condition for their entanglement. This criterion has been used to characterize the strength of entanglement in several previous experiments [8–11].

It is relatively easy to show that for pure input squeezed states ($\{\Delta^2 \hat{X}^+_{\rm sqz,1} \cdot \Delta^2 \hat{X}^-_{\rm sqz,1}, \Delta^2 \hat{X}^+_{\rm sqz,2} \cdot \Delta^2 \hat{X}^-_{\rm sqz,2}\} = 1$) and equal optical loss for beams $x$ and $y$, the dependence of $\mathcal{E}$ on detection efficiency is given by

$$\mathcal{E} = 4\left(1 - \eta + \frac{2\eta - 1}{\eta(\Delta^2 \hat{X}^+_{\rm sqz,ave} + 1/\Delta^2 \hat{X}^+_{\rm sqz,ave} - 2) + 2}\right)^2 \tag{31}$$

Notice that when $\eta = 0.5$, $\mathcal{E} = 1$, independent of the level of squeezing. This defines a boundary such that if $\eta > 0.5$ the EPR paradox criterion is satisfied for any level of squeezing, and if $\eta < 0.5$ it can never be satisfied. This is a striking contrast to the inseparability criterion which, as we showed earlier, is satisfied for any level of squeezing and any detection efficiency. The reason for this difference is that the inseparability criterion is independent of the purity of the entanglement (ie. independent of $\Delta^2 \hat{X}^+_{\rm sqz,1} \cdot \Delta^2 \hat{X}^-_{\rm sqz,1}$ and $\Delta^2 \hat{X}^+_{\rm sqz,2} \cdot \Delta^2 \hat{X}^-_{\rm sqz,2}$), a property that the EPR paradox criterion is very sensitive to. Optical loss changes the purity of the entanglement and therefore effects the EPR paradox and inseparability criteria differently. However, if $\eta = 1$ the measured entangled state is pure, and both criteria are monotonically increasing functions of $\Delta^2 \hat{X}^+_{\rm sqz,ave}$ in the range $0 < \Delta^2 \hat{X}^+_{\rm sqz,ave} < 1$, with $\mathcal{E} = \mathcal{I} = 1$ at $\Delta^2 \hat{X}^+_{\rm sqz,ave} = 1$. Therefore, in the limit of pure measured entanglement, the inseparability and EPR paradox criteria become qualitatively equivalent.

### D The photon number diagram

Applications have been proposed for quadrature entanglement in the field of quantum information [23, 24]. For almost all of these applications, a pure entangled state is desired [25]. Due to the unavoidable losses in any real system however, a perfectly pure entangled state is unachievable. It is therefore essential to characterize the effect of impurity on the outcome of any application of entanglement. We have seen already, that impurity has different effects on the degrees of inseparability and EPR paradox. It may not be such a surprise therefore, that the effect of impurity varies from application to application.

To illustrate the point we consider two well known potential applications related to quantum information, unity gain quantum teleportation [16, 26–28] and dense coding [5, 29]. We analyze the performance of these applications as a function of the purity of the entanglement, and its strength inferred from the inseparability criterion.

A nice feature of some discrete variable measures of an entanglement resource, such as Von Neumann entropy[19] and relative entropy[30], is that they vary proportionally with the size of the resource. That is, if the number of entangled photon pairs doubles the value of the measure doubles. This is not the case for the inseparability criterion. In fact, as the strength of the entanglement increases, the inseparability criterion approaches zero. Alternatively, in a manner analogous to discrete variable entanglement measures, we can examine the average number of photons per bandwidth per time required to generate the entanglement resource [18]. The average number of photons per bandwidth per time in the sideband $\omega$ of an optical beam is given by

$$\begin{aligned}\bar{n}(\omega) &= \langle \hat{a}^\dagger(\omega)\hat{a}(\omega)\rangle \\ &= \frac{1}{4}\Big\langle\big(\hat{X}^+ - i\hat{X}^-\big)\big(\hat{X}^+ + i\hat{X}^-\big)\Big\rangle \\ &= \frac{1}{4}\Big(\Big\langle(\hat{X}^+)^2\Big\rangle+\Big\langle(\hat{X}^-)^2\Big\rangle+i\left\langle\hat{X}^+\hat{X}^- - \hat{X}^-\hat{X}^+\right\rangle\Big) \\ &= \left|\alpha^+\right|^2 + \left|\alpha^-\right|^2 + \frac{1}{4}\Big(\Delta^2\hat{X}^+ + \Delta^2\hat{X}^- - 2\Big). \quad (32)\end{aligned}$$

We see that with only vacuum in the sideband $\Delta^2\hat{X}^+ = \Delta^2\hat{X}^- = 1$ and $\alpha^\pm = 0$, so no photons are present. If the state is squeezed, however, then $\Delta^2\hat{X}^+ + \Delta^2\hat{X}^- > 2$ always, and therefore $\bar{n} > 0$. As the squeezing improves the average number of photons in the state increases. Since entanglement may be generated by interfering a pair of squeezed beams we can see that to maintain an entangled resource of a given strength (or a given $\mathcal{I}$) will also require some non-zero average number of photons. The mean number of photons in an entangled state $\bar{n}_{\rm total}$ is just the sum of the number in beams $x$ and $y$

$$\begin{aligned}\bar{n}_{\rm total} &= \bar{n}_x + \bar{n}_y \quad (33)\\ &= \frac{1}{4}\Big(\Delta^2\hat{X}_x^+ + \Delta^2\hat{X}_x^- + \Delta^2\hat{X}_y^+ + \Delta^2\hat{X}_y^-\Big) - 1, \quad (34)\end{aligned}$$

where since the coherent amplitudes $\alpha_x^\pm$ and $\alpha_y^\pm$ have no relevance to the correlation matrix characterising our entanglement, and are easily accounted for, we have neglect contributions from them setting $\alpha_x^\pm = \alpha_y^\pm = 0$. As stated earlier, some fraction of $\bar{n}_{\rm total}$ is required to maintain the strength of the entanglement. A contribution is also made by the impurity of the squeezed beams used to generate the entanglement; and by the decoherence experienced by the state after production. Of course, the photons in a quadrature entangled state are indistinguishable from one another so that a definite separation of photons into distinct categories is not possible. This separation is possible however, when only the average number of photons within a quadrature entangled state per bandwidth per time is considered. The strength of the entanglement ($\mathcal{I}$) dictates a minimum average number of photons $\bar{n}_{\rm min}$ per bandwidth per time that are required to maintain the entanglement. The remaining photons can (on average) be separated into photons that are present due to bias between the amplitude and phase quadratures of the entangled beams $\bar{n}_{\rm bias}$, and excess photons that are the result of the impurity of the entanglement $\bar{n}_{\rm excess}$.

For entanglement that is symmetric between beams $x$ and $y$ such as is analyzed in this paper, the average number of excess photons per bandwidth per time $\bar{n}_{\rm excess}$ can be found by considering the lossless interference of the two entangled beams in phase on a 50/50 beam splitter. In this case the output beams (labelled with the sub-scripts 'out1' and 'out2' here) would exhibit squeezing with squeezed quadrature variances of $\Delta^2\hat{X}^+_{\rm sqz,out1} = \Delta^2\hat{X}^+_{x\pm y}$ and $\Delta^2\hat{X}^-_{\rm sqz,out2} = \Delta^2\hat{X}^-_{x\pm y}$, respectively. From eq. (25) we see that the strength of our entanglement, $\mathcal{I}$, depends only on the squeezing of these output beams. Any impurity in the entanglement causes the output beams to be non-minimum uncertainty $(\{\Delta^2\hat{X}^+_{\rm sqz,out1}\Delta^2\hat{X}^-_{\rm sqz,out1}, \Delta^2\hat{X}^+_{\rm sqz,out2}\Delta^2\hat{X}^-_{\rm sqz,out2}\} > 1)$. To determine the average number of photons in the entangled state due to impurity, $\bar{n}_{\rm excess}$, we can simply compare the mean number of photons in the entangled state $\bar{n}_{\rm total}$ to the number that would be in the state if it was perfectly pure, $\bar{n}_{\rm pure}$,

$$\begin{aligned}\bar{n}_{\rm excess} &= \bar{n}_{\rm total} - \bar{n}_{\rm pure} \quad (35)\\ &= \bar{n}_x + \bar{n}_y - \bar{n}_{\rm pure}. \quad (36)\end{aligned}$$

$\bar{n}_{\rm pure}$ can be thought of as the average number of photons per bandwidth per time required to generate two pure squeezed beams with the same level of squeezing as the two output beams. When $\mathcal{I} \geq 1$ no entanglement is present between beams $x$ and $y$, and no squeezing is required. We therefore find $\bar{n}_{\rm min} = 0$ and $\bar{n}_{\rm total} = \bar{n}_{\rm excess} + \bar{n}_{\rm bias}$. For the remainder of this paper we only consider the more interesting situation when entanglement is present, restricting ourselves to $\mathcal{I} < 1$. In this case since the two output beams have squeezed quadrature variances of $\Delta^2\hat{X}^+_{x\pm y}$ and $\Delta^2\hat{X}^-_{x\pm y}$ respectively, $\bar{n}_{\rm pure}$ is given by

$$\bar{n}_{\rm pure} = \frac{1}{4}\left(\Delta^2\hat{X}^+_{x\pm y} + \frac{1}{\Delta^2\hat{X}^+_{x\pm y}} + \Delta^2\hat{X}^-_{x\pm y} + \frac{1}{\Delta^2\hat{X}^-_{x\pm y}}\right) - 1. \quad (37)$$

$\bar{n}_{\rm excess}$ can then be directly obtained from eq. (36).

$\bar{n}_{\rm pure}$ can be separated into a component due to bias in the entanglement $\bar{n}_{\rm bias}$ and a component required to maintain the entanglement $\bar{n}_{\rm min}$

$$\bar{n}_{\rm pure} = \bar{n}_{\rm min} + \bar{n}_{\rm bias}. \quad (38)$$

$\bar{n}_{\rm min}$ is directly dependent on the strength of the entanglement $\mathcal{I}$, and is therefore independent of local reversible operations performed individually on beams $x$ and $y$. The photons resulting from bias between the amplitude and phase quadratures of the entangled state, however, may be completely eliminated by performing equal local squeezing operations on beams $x$ and $y$ [18]. After performing these operations $\bar{n}_{\rm pure}$ becomes

$$\bar{n}_{\rm pure}^{'} = \frac{1}{4}\left(g^2\Delta^2\hat{X}^{+}_{x\pm y} + \frac{1}{g^2\Delta^2\hat{X}^{+}_{x\pm y}} + \frac{\Delta^2\hat{X}^{-}_{x\pm y}}{g^2} + \frac{g^2}{\Delta^2\hat{X}^{-}_{x\pm y}}\right) - 1, \tag{39}$$

where $g$ is the gain of the squeezing operations. It is relatively easy to show that $\bar{n}_{\rm pure}^{'}$ is minimized, and therefore $\bar{n}_{\rm bias}$ is eliminated, when $g^2 = \sqrt{\Delta^2\hat{X}^{-}_{x\pm y}/\Delta^2\hat{X}^{+}_{x\pm y}}$, and we find that

$$\begin{aligned}\bar{n}_{\rm min} &= \frac{1}{2}\left(\sqrt{\Delta^2\hat{X}^{+}_{x\pm y}\Delta^2\hat{X}^{-}_{x\pm y}} + \frac{1}{\sqrt{\Delta^2\hat{X}^{+}_{x\pm y}\Delta^2\hat{X}^{-}_{x\pm y}}}\right) - 1 \\ &= \frac{1}{2}\left(\mathcal{I} + \frac{1}{\mathcal{I}}\right) - 1, \end{aligned} \tag{40}$$

where $\bar{n}_{\rm min}$ is the minimum mean number of photons per bandwidth per time required to generate entanglement of a given strength $\mathcal{I}$. We see that $\bar{n}_{\rm min}$ is completely determined by $\mathcal{I}$ and is monotonically increasing as $\mathcal{I} \to 0$. The average number of photons present in the entanglement per bandwidth per time as a result of bias can then also be deterimined $\bar{n}_{\rm bias} = \bar{n}_{\rm pure} - \bar{n}_{\rm min}$.

We can now separate the average photon number per bandwidth per time in a quadrature entangled state into three categories; photons required to maintain the entanglement $\bar{n}_{\rm min}$, photons produced by bias between the amplitude and phase quadratures $\bar{n}_{\rm bias}$, and excess photons resulting from impurity $\bar{n}_{\rm excess}$. All three average photon numbers can be calculated from measurements of $\Delta^2\hat{X}^{\pm}_{x}$, $\Delta^2\hat{X}^{\pm}_{y}$, and $\Delta^2\hat{X}^{\pm}_{x\pm y}$. An entangled state can then be conveniently and intuitively analyzed on a three dimensional diagram as shown in fig. 1, with $\bar{n}_{\rm min}$, $\bar{n}_{\rm bias}$, and $\bar{n}_{\rm excess}$, forming each of the axes. Note that, in a manner analogous to that performed for $\bar{n}_{\rm pure}$ above, $\bar{n}_{\rm excess}$ may be broken into two parts: the average number of photons required to produce the impurity of the entanglement, and the average number of photons generated by bias between the amplitude and phase quadratures caused by the impurity of the state. We do not perform this separation explicitly here, since the exact distribution of excess photons is of much less significance than that for the photons necessary to generate the entanglement. Including this extra parameter, and assuming the entanglement is of the same form as is discussed earlier, the correlation matrix of section III A can be fully characterized by these photon number parameters.

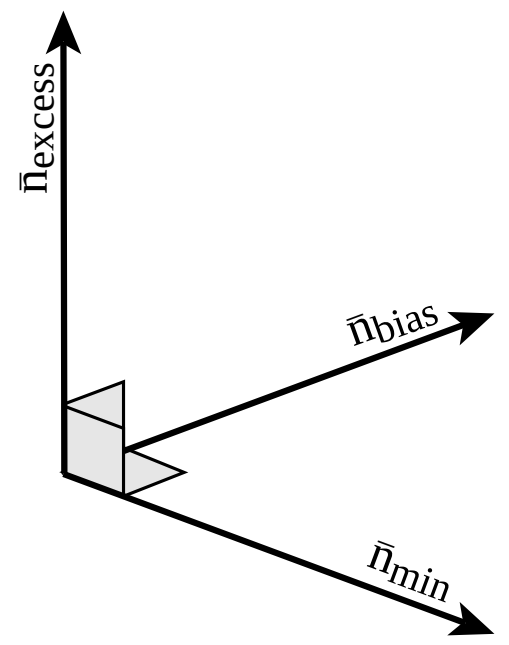

FIG. 1: An entangled state can be represented on a three dimensional photon number diagram.

An analogy can be made between the $\bar{n}_{\rm min}$-$\bar{n}_{\rm excess}$ plane of the photon number diagram and the tangle/linear entropy analysis often performed for discrete variable entanglement[31]. In both cases the entanglement is represented on a plane with one axis representing the strength of the entanglement ($\bar{n}_{\rm min}$ for continuous variables, and the tangle for discrete variables), and the other axis representing the purity of the state ($\bar{n}_{\rm excess}$ for continuous variables and the linear entropy for discrete variables). Unlike the discrete variable case where the region of the tangle-linear entropy plane occupied by physical states is bounded, the set of continuous variable entangled states spans the entire $\bar{n}_{\rm min}$-$\bar{n}_{\rm excess}$ plane. The difference occurs because the discrete quantum states analyzed on the tangle-linear entropy plane involve a finite and fixed number of photons. This restriction limits both the strength of the entanglement (the tangle) and the purity (the linear entropy). Continuous variable entangled states have no such limitation.

It is interesting to consider whether $\bar{n}_{\rm min}$ is a good measure of entanglement. Formally, a good measure of the entanglement of the state $\rho$, $E(\rho)$, must satisfy the following criteria[19, 20]:

I. $E(\rho) = 0$ if and only if $\rho$ is separable,

II. $E(\rho)$ is left invariant under local unitary operations,

III. $E(\rho)$ is non-increasing under local general measurements and classical communication,

IV. Given two separate entangled states $\rho_1$ and $\rho_2$ such that $\rho = \rho_1 \otimes \rho_2$, $E(\rho) = E(\rho_1) + E(\rho_2)$.

Duan *et al.* demonstrated that $\mathcal{I} = 1$ if and only if the state under interrogation is separable. It is clear then that $\bar{n}_{\rm min} = 0$ if and only if the state under interrogation is separable, and therefore criterion I is true for $\bar{n}_{\rm min}$. Furthermore, since characterization of $\mathcal{I}$ requires that the states correlation matrix be taken into a standard form, both $\mathcal{I}$ and $\bar{n}_{\rm min}$ are invariant under local unitary operations so that criterion II is true. As yet we have no conclusion about the validity of criterion III for $\bar{n}_{\rm min}$. It seems likely that it is valid since an increase in $\bar{n}_{\rm min}$ is equivalent to an increase in the quantum correlation between fields $x$ and $y$, which should not be possible through local general measurements and classical communication. Finally, given two separate entangled states the minimum average number of photons per bandwidth per time required to generate both states is simply the sum of the minimum average number of photons per bandwidth per time required to generate each state, $\bar{n}_{\rm min} = \bar{n}_{{\rm min},1} + \bar{n}_{{\rm min},2}$, so that criterion IV is valid. We see therefore that $\bar{n}_{\rm min}$ satisfies three of the four criteria for a good entanglement measure, and although we have not shown so here, we believe it is likely to satisfy the remaining criterion. $\bar{n}_{\rm min}$ is a particularly elegant measure of entanglement due to its physical significance.

### 1 Entanglement criteria and the photon number diagram

We can represent the inseparability and EPR paradox criteria on the photon number diagram. As can be seen from eq. (40), for entanglement symmetric between beams $x$ and $y$ the degree inseparability can be expressed solely as a function of $\bar{n}_{\rm min}$

$$\mathcal{I} = \bar{n}_{\rm min} + 1 - \sqrt{(\bar{n}_{\rm min}+1)^2 - 1}. \tag{41}$$

The same is not true for the EPR paradox criterion. This result is unsurprising, we have already found that the EPR paradox is sensitive to the impurity of the entangled state which can be expressed in terms of $\bar{n}_{\rm excess}$. The degree of EPR paradox can be obtained from the amplitude and phase quadrature conditional variances between beams $x$ and $y$ (see eq. (27)). We see from eq. (28) that the amplitude and phase quadrature conditional variances are defined by $\Delta^2\hat{X}_x^{\pm}$, $\Delta^2\hat{X}_y^{\pm}$, and $\left|\langle \delta X_x^{\pm}\delta X_y^{\pm}\rangle\right|$. For simplicity here we assume the entanglement is symmetric between amplitude and phase quadratures. This assumption is true for the entanglement analyzed in this paper at sideband frequencies above around 5 MHz, and has the consequence that there are no photons in the entangled state due to bias $\bar{n}_{\rm bias} = 0$. We then find that

$$\Delta^2\hat{X} = \Delta^2\hat{X}_x^{+} = \Delta^2\hat{X}_x^{-} = \Delta^2\hat{X}_y^{+} = \Delta^2\hat{X}_y^{-} = \bar{n}_{\rm total}+1, \tag{42}$$

and can express $\left|\langle \delta X_x^{\pm}\delta X_y^{\pm}\rangle\right|$ in terms of $\bar{n}_{\rm min}$ and $\bar{n}_{\rm excess}$ as

$$\left|\langle \delta X_x^{\pm}\delta X_y^{\pm}\rangle\right| = \bar{n}_{\rm excess} + \sqrt{(\bar{n}_{\rm min})^2 - 1}. \tag{43}$$

The degree of EPR paradox can then also be written in terms of $\bar{n}_{\rm min}$ and $\bar{n}_{\rm excess}$

$$\mathcal{E} = \left(\frac{2\bar{n}_{\rm excess}\left(\bar{n}_{\rm min}+1-\sqrt{(\bar{n}_{\rm min}+1)^2-1}\right)+1}{\bar{n}_{\rm excess}+\bar{n}_{\rm min}+1}\right)^2. \tag{44}$$

Since we have assumed that $\bar{n}_{\rm bias} = 0$, the degree of EPR paradox can be represented as contours on the $\bar{n}_{\rm min}$-$\bar{n}_{\rm excess}$ plane of the photon number diagram. This representation is shown in fig. 15 a), the curvature of the contours demonstrates again the sensitivity of the EPR paradox to impurity.

It is interesting to note that in the extrema of $\bar{n}_{\rm excess} \to 0$ and $\bar{n}_{\rm excess} \to \infty$, the degree of EPR paradox becomes a function of only $\bar{n}_{\rm min}$, and can be written in terms of the degree of inseparability as

$$\mathcal{E}_{\bar{n}_{\rm excess}\to 0} = \frac{4\mathcal{I}^2}{(\mathcal{I}^2+1)^2} \tag{45}$$

$$\mathcal{E}_{\bar{n}_{\rm excess}\to \infty} = 4\mathcal{I}^2. \tag{46}$$

We see again that for pure entanglement ($\bar{n}_{\rm excess} = 0$) $\mathcal{I} < 1$ implies $\mathcal{E} < 1$. In contrast, for extremely impure entanglement ($\bar{n}_{\rm excess} \to \infty$), we see that to observe the EPR paradox requires $\mathcal{I} < 0.5$. This result has the consequence that if the squeezed beams used to generate the entanglement have squeezed variances $\{\Delta^2\hat{X}^{+}_{\rm sqz,1}, \Delta^2\hat{X}^{+}_{\rm sqz,2}\} < 0.5$, then no matter how large the anti-squeezed variances, the EPR paradox can be demonstrated.

### 2 Quantum teleportation and the photon number diagram

Quantum information protocols are also representable on the photon number diagram. In this paper we consider two well-known examples, quantum teleportation and dense coding.

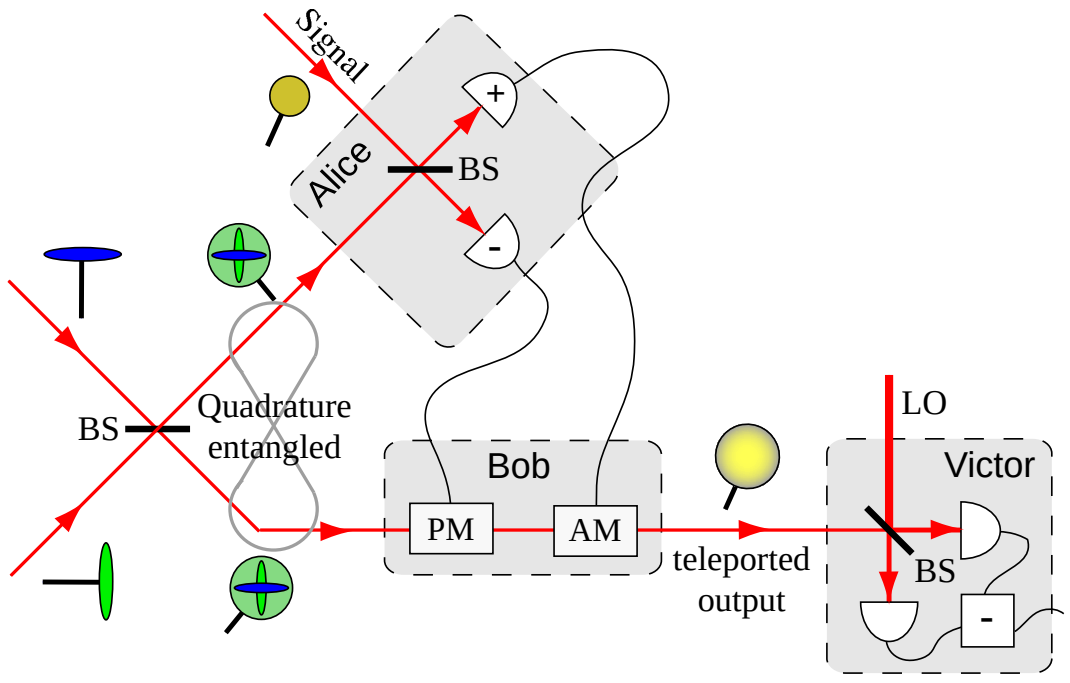

FIG. 2: Schematic of a quantum teleportation experiment, detectors labelled with the symbols + and - are amplitude and phase detectors respectively, BS: beam splitter, AM: amplitude modulator, PM: phase modulator, LO: local oscillator.

The uncertainty principle of quantum mechanics fundamentally limits both the ability to measure and to reconstruct quantum states. Since teleportation requires both measurement of the original state, and then reconstruction at a distant location, it was therefore thought that teleportation was also fundamentally limited by the uncertainty principle. In 1993, however, Bennett *et. al* [3] discovered that by using entangled photon pairs in the measurement and reconstruction processes perfect teleportation could be facilitated. Their proposal has been generalized to the continuous variable regime [26, 28], and a schematic of the continuous variable scheme is shown in fig. 2. A number of methods exist to characterize the success of continuous variable teleportation (for a summary see [33]), in this paper we consider the most well known measure, the fidelity of teleportation. Fidelity measures the state-overlap between the teleporter input $|\psi_{\rm in}\rangle$ and output $\hat{\rho}_{\rm out}$ states, and is given by

$$\mathcal{F} = \langle\psi_{\rm in}|\hat{\rho}_{\rm out}|\psi_{\rm in}\rangle. \tag{47}$$

$\mathcal{F} = 1$ implies perfect overlap between the input and output states and therefore perfect teleportation, without entanglement the fidelity is limited to $\mathcal{F} \leq 0.5$, and $\mathcal{F} = 0$ if the input and output states are orthogonal. Again assuming that the entanglement is unbiased ($\bar{n}_{\rm bias} = 0$), the fidelity of unity gain coherent state teleportation using quadrature entangle-ment [26, 28] may be expressed as

$$\mathcal{F} = \frac{1}{1+\mathcal{I}}. \tag{48}$$

We see that the success of the teleportation protocol depends only on the degree of inseparability. This results in vertical efficacy contours for teleportation when represented on the photon number diagram, as can be seen in fig. 15 b). The shading in fig. 15 b) indicates the area of the photon number diagram for which the more stringent *no cloning* teleportation limit is not satisfied [32]. Note that, if the teleportation protocol was operated at non-unity gain, the protocol would become sensitive to impurity and the teleportation efficacy contours would be curved. Although the non-unity gain regime is significant for quantum information protocols such as optimum entanglement swapping [33], we will not consider it here.

### 3 *Dense coding and the photon number diagram*

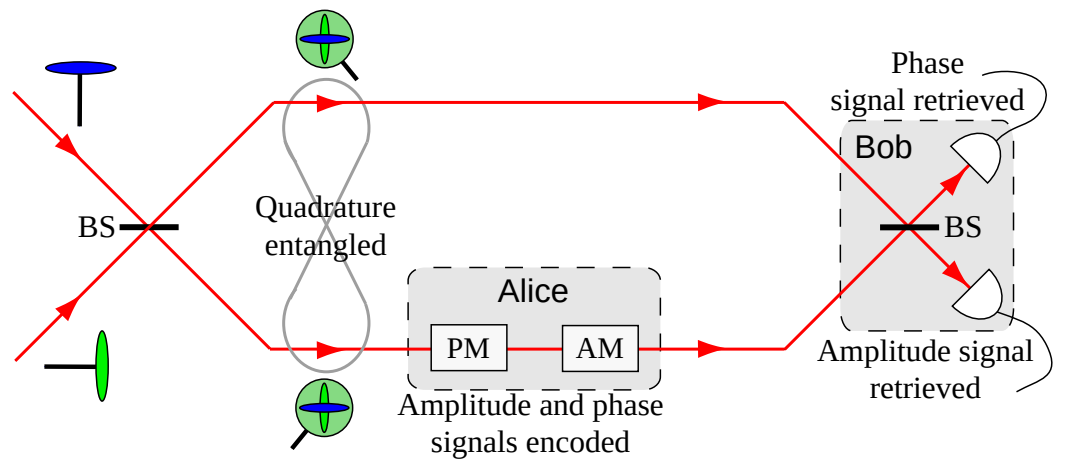

FIG. 3: Schematic of a dense coding experiment, BS: beam splitter, AM: amplitude modulator, PM: phase modulator.

Dense coding was first proposed by Bennett *et. al* [34] in 1992, when they showed that by utilizing shared entanglement between the sending (Alice) and receiving (Bob) stations, a single communication channel can achieve a higher information transfer rate than is physically possible using the same resources (i.e. the same number of photons) but without entanglement.

An upper bound to the information transfer rate of a bandwidth limited Gaussian information channel is given by the Shannon capacity $C$ [35]

$$C = \frac{\log_2(1+R)}{2}, \tag{49}$$

where $R = \Delta^2\hat{S}/\Delta^2\hat{N}$ is the signal to noise ratio of the channel, with $\Delta^2\hat{S}$ and $\Delta^2\hat{N}$ being the variance of the signal and noise respectively. Dense coding in the continuous variable regime was first proposed by Braunstein and Kimble in 2000 [29], and a detailed discussion may be found in [5]. A schematic diagram of the proposal of Braunstein and Kimble is given in fig. 3.

In this paper we restrict ourselves to the comparison of the channel capacities achievable using a squeezed state and using a dense coding protocol based on quadrature entanglement. To obtain a fair comparison of the two schemes we define the total average number of photons allowed in the beam encoded with information $\bar{n}_{\rm encoding}$. In both the squeezed state and entangled state based dense coding schemes some of these photons must be used to generate the quantum state, and the remaining photons can be used to encode signals. For the squeezed state scheme the number of photons in the squeezed state is given by

$$\bar{n}_{\rm sqz} = \frac{1}{4}\left(\Delta^2\hat{X}_{\rm sqz} + \frac{1}{\Delta^2\hat{X}_{\rm sqz}} - 2\right), \tag{50}$$

where $\Delta^2\hat{X}_{\rm sqz}$ is the variance of the squeezed quadrature. The remaining $\bar{n}_{\rm encoding} - \bar{n}_{\rm sqz}$ photons are used to encode signals on the squeezed quadrature of the beam. This results in a channel with signal variance given by $\Delta^2\hat{S}_{\rm sqz} = 4(\bar{n}_{\rm encoding} - \bar{n}_{\rm sqz})$, and noise variance given by $\Delta^2\hat{N}_{\rm sqz} = \Delta^2\hat{X}_{\rm sqz}$. The squeezed state channel capacity is then

$$C_{\rm sqz} = \log_2\left(1 + \frac{4(\bar{n}_{\rm encoding} - \bar{n}_{\rm sqz})}{\Delta^2\hat{X}_{\rm sqz}}\right). \tag{51}$$

Optimizing the ratio of the mean number of photons per bandwidth per time used to generate squeezing and the mean number of photons per bandwidth per time used to encode the signal we arrive at the optimum squeezed state channel capacity [5]

$$C_{\rm sqz,opt} = \log_2(1 + 2\bar{n}_{\rm encoding}). \tag{52}$$

Let us now consider the dense coding scheme. Again, we make the assumption that the entanglement is symmetric between the amplitude and phase quadratures. In this case we can use the amplitude and phase quadratures as independent channels, and find that the noise variance of each channel is given by $\Delta^2\hat{N}_{\rm EPR} = \mathcal{I} = \Delta^2\hat{X}^+_{x\pm y} = \Delta^2\hat{X}^-_{x\pm y}$. $\bar{n}_{\rm total}$ as defined previously is the average number of photons per bandwidth per time in the entangled state before encoding of any signals. These photons are split evenly between the two entangled beams, therefore on average $\bar{n}_{\rm encoding} - \bar{n}_{\rm total}/2$ photons per bandwidth per time are available for encoding. The amplitude and phase quadrature signal variances are both then given by $\Delta^2\hat{S}_{\rm EPR} = \bar{n}_{\rm encoding} - \bar{n}_{\rm total}/2$, which is attenuated by a factor of four when compared to the squeezed state signal variance. This attenuation is the result of two effects, a factor of two arises because the signal photons must be shared between the amplitude and phase quadratures of the entangled beam, and another factor of two is due to the 50/50 beam splitter required before measurement. We then obtain the entangled state channel capacity

$$C_{\rm EPR} = \log_2\left(1 + \frac{\Delta^2\hat{S}_{\rm EPR}}{\mathcal{I}}\right) \tag{53}$$

$$= \log_2\left(1 + \frac{\bar{n}_{\rm encoding} - (\bar{n}_{\rm min} + \bar{n}_{\rm excess})/2}{\bar{n}_{\rm min} + 1 - \sqrt{(\bar{n}_{\rm min}+1)^2 - 1}}\right) \tag{54}$$

When the average number of photons available to the dense coding protocol is large ($\bar{n}_{\rm encoding} \to \infty$), the dense coding channel capacity becomes independent of the number of photons present due to impurity in the entanglement. This is shown in fig. 15 d) which plots contours of the ratio $C_{\rm EPR}/C_{\rm sqz}$ for large $\bar{n}_{\rm encoding}$. We see that in this limit the dense coding channel capacity exceeds the optimum achievable squeezed state channel capacity for $\bar{n}_{\rm min} > 0.25$. When the average number of photons available to the dense coding protocol is small however, the dense coding channel capacity can be extremely sensitive to impurity. This is perhaps not a surprise, since every photon that exists in the entangled state is one less that may be used to encode signals. Clearly, in the limit that $\bar{n}_{\rm encoding} = (\bar{n}_{\rm min} + \bar{n}_{\rm excess})/2$, no photons remain to encode signals, and therefore $C_{\rm EPR} = 0$. The ratio $C_{\rm EPR}/C_{\rm sqz}$ for small $\bar{n}_{\rm encoding}$ is shown as a function of $\bar{n}_{\rm min}$ and $\bar{n}_{\rm excess}$ in fig. 15 c), and indeed the contours are strongly curved.

## IV. EXPERIMENT

The previous section described methods presently available to characterize continuous variable entangled states. In particular we discussed the correlation matrix which can be used to fully characterize Gaussian entanglement, the inseparability and EPR paradox criteria, and a new representation the photon number diagram. In this section we describe the methods used in our experiment to generate a pair of entangled beams. We then present experimental results for each entanglement characterization technique over the frequency range from 2.5 to 10 MHz. We examine the effect of loss on the inseparability and EPR paradox criteria demonstrating qualitative differences, and use the photon number diagram to predict the efficacy of our entanglement in the quantum information protocols introduced earlier.

### A Generation of quadrature squeezing

The laser source for our experiment was a 1.5 W monolithic non-planar ring Nd:YAG laser at 1064 nm. Its output was split into two beams as shown in fig. 4, one of these beams was mode-matched into a second harmonic generator (SHG) to produce 532 nm light to pump a pair of optical parametric amplifiers (OPAs); and the other was used to seed the OPAs and for homodyne detection of our entangled beams. The SHG consisted of a 7.5 mm long hemi-lithic MgO:LiNbO$_3$ crystal and an output coupler. One end of the MgO:LiNbO$_3$ crystal had a 10 mm radius of curvature and was coated for high reflection at 1064 and 532 nm. The other end was flat and anti-reflection coated at both 1064 and 532 nm. The output coupler had a radius of curvature of 25 mm, it was anti-reflection coated for 532 nm ($R_{532} \approx 7$ %), and had 92 % reflection of 1064 nm. 23 mm separated the MgO:LiNbO$_3$ crystal and the output coupler, this created a cavity mode for the resonant 1064 nm light with a 27 $\mu$m waist at the center of the MgO:LiNbO$_3$ crystal. A 29.7 MHz electro-optic modulation was applied to the MgO:LiNbO$_3$ crystal, detecting and de-modulating the transmitted light intensity at 29.7 MHz provided a Pound-Drever-Hall (PDH) type error signal [36] which was then used to control the length of the SHG resonator. The SHG provided 370 mW of 532 nm light with 50 % conversion efficiency.

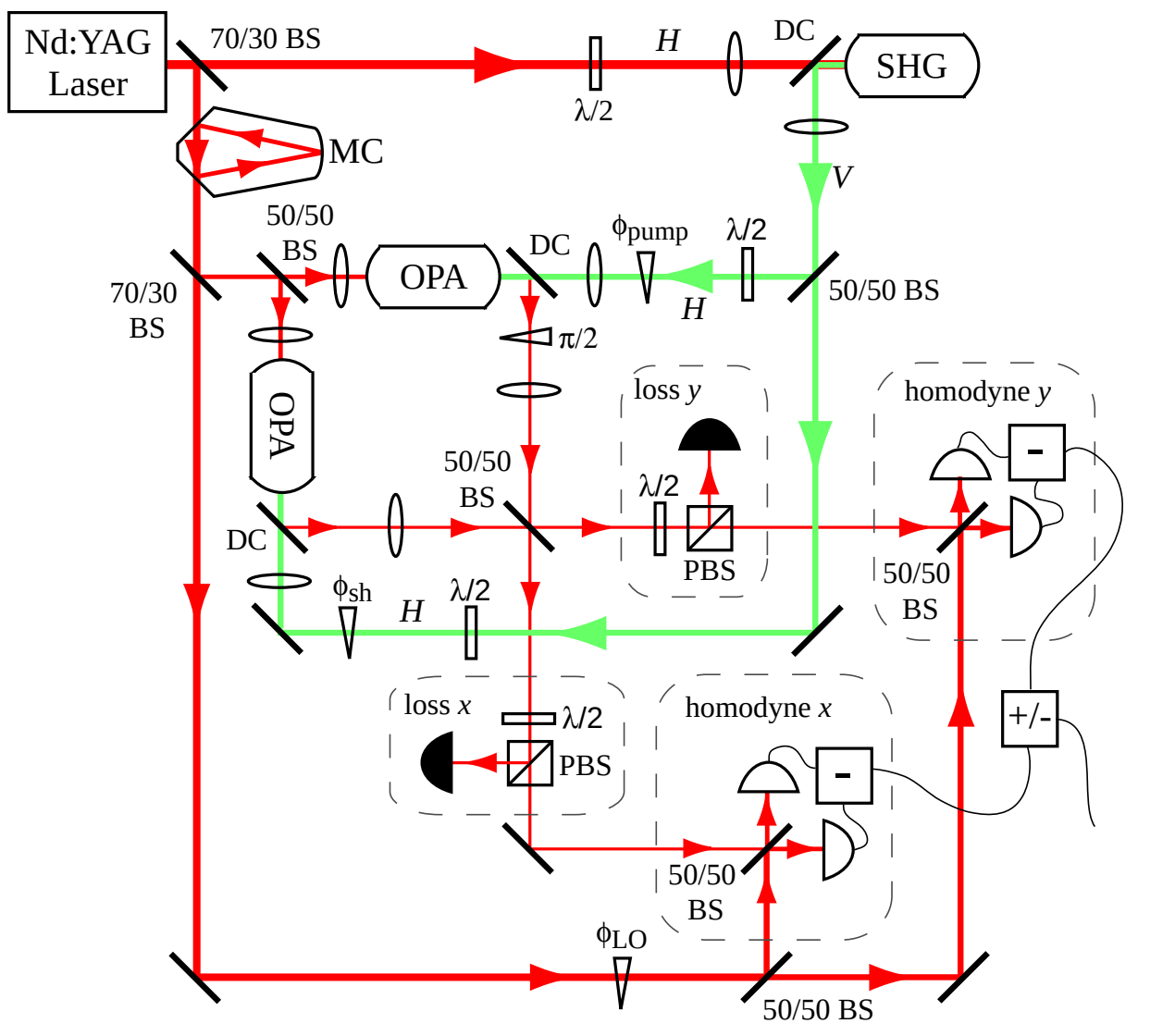

FIG. 4: Experimental schematic. $x$ and $y$ respectively label the entangled beams, BS (PBS): 50/50 (polarizing) beam splitter, $\lambda/2$: half-wave plate, $\phi$ and $\theta$: phase shift.

The remaining 1064 nm beam was transmitted through a high finesse ring cavity to reduce its spectral noise. This cavity was based on a LIGO advanced gravitational wave mode cleaner design [37]. It consisted of two closely spaced flat 45° angled input/output coupling mirrors, and a 1 m radius of curvature mirror coated for high reflection at normal incidence, and had a total cavity length of roughly 50 cm. All three mirrors were coated by Research-Electro-Optics (REO) with part-per-million tolerances. Since the reflectivity of the angled input/output couplers depended on the polarization of the input field, the mode cleaner had two modes of operation, high finesse and low finesse, which had approximate finesses of 2000 and 170, and corresponding linewidths of 300 kHz and 3 MHz, respectively. Above these linewidths spectral noise from the laser is significantly attenuated on transmission. In our experiment we utilized the low finesse mode to maximize the power transmitted through the cavity, we found that the output was quantum noise limited at 6 MHz. The laser frequency was locked to the mode cleaner using tilt locking [38], a phase sensitive spatial mode interference technique analogous to PDH locking. Unlike PDH locking this technique introduces no modulation sidebands, an advantage in our case since modulation sidebands can transfer power into the squeezing spectrum produced by our OPAs.

The mode cleaner output beam was split to provide seeds for our two OPAs, as well as homodyne local oscillators for interrogation of the two entangled beams. The OPAs were iden-tical in design to the SHG, except that the output coupling mirrors were 96 % reflective at 1064 nm. They were each seeded through the high reflective surface of the MgO:LiNbO$_3$ crystal. A 30.5 MHz electro-optic modulation was applied to each crystal which allowed the length of both OPA resonators to be actively controlled. The 532 nm light was split into two parts and used to pump the OPAs. This results in either amplification or deamplification of the seed, depending on the relative phase between the pump and seed. The 29.7 MHz modulation on the SHG crystal produced a 29.7 MHz phase modulation on both 532 nm pump beams. This caused a modulation of the amplification of the OPAs that could be used to control the relative phase between the pump and seed. By detecting the reflected light from each OPA, and de-modulating at 29.7 MHz we generated error signals to lock each OPA to either amplification or deamplification. When locked to amplification, the 1064 nm output exhibited phase squeezing, and when locked to de-amplification it exhibited amplitude squeezing. Pick-up across the copper plates used to electro-optically modulate our OPAs couples noise directly into the phase quadrature of the output beams. We therefore chose to lock to amplitude squeezing. We balanced the power in the squeezed beams by adjusting the OPA seed powers and analyzed the squeezing using homodyne detection with roughly 84 % efficiency. The homodyne detector could be locked to detect either the amplitude or phase quadrature of the input beam. Throughout this paper, locking to the amplitude quadrature was enabled through a phase modulation on the input beam, and locking to the phase quadrature was achieved when the power splitting within the detector was balanced. All of the spectra presented in this paper were obtain from homodyne detector output photo-currents analyzed in a Hewlett-Packard E4405B spectrum analyzer with 300 kHz resolution bandwidth and 300 Hz video bandwidth over the frequency range from 2.5 to 10 MHz. Each spectra was at least 4.5 dB above the detection darknoise which was taken into account. Typical amplitude squeezing spectra for each of our OPAs are shown in fig. 5. The OPAs produced near identical spectra with an optimum of 3.7 dB of squeezing at 6.5 MHz. Both spectra are degraded at low frequencies due to the resonant relaxation oscillation of our laser, and at high frequencies due to the bandwidth of the OPAs.

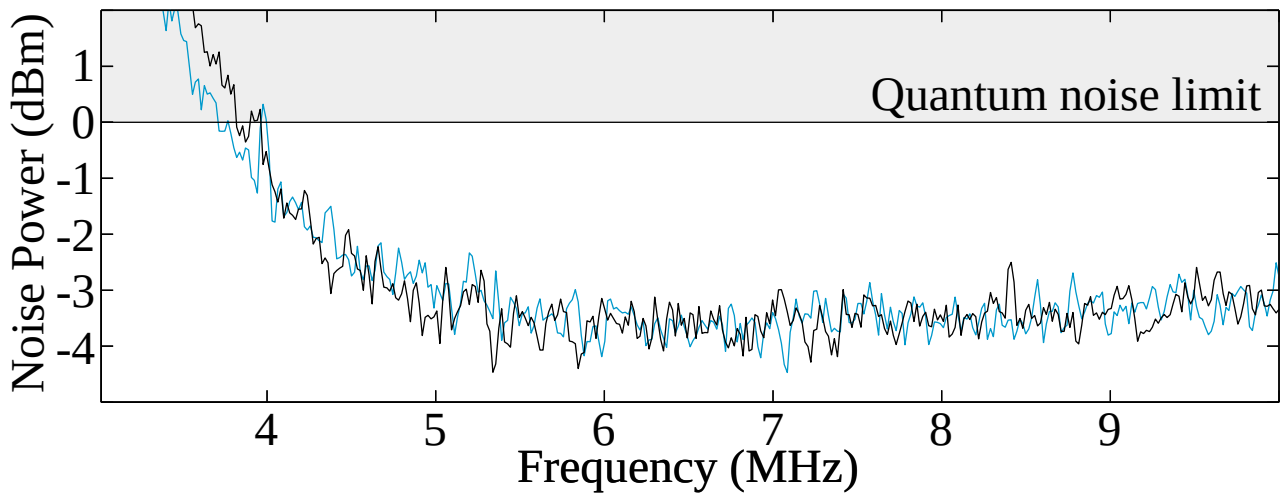

FIG. 5: Squeezing spectra observed from the two OPAs, normalized to the quantum noise limit.

### B Generation and measurement of entanglement

We generated quadrature entanglement by combining our two amplitude squeezed beams with relative phase of $\pi/2$ on a 50/50 beam splitter as discussed in section II. A visibility of $98.7\pm0.3$ % was observed for the process, and the relative phase was controlled at $\pi/2$ by actively balancing the power in the two entangled beams. Each entangled beam was interrogated in a balanced homodyne detector that could be locked to detect either its phase or amplitude quadrature. The efficiency of the detection process was approximately 86 %, with loss contributed equally by the homodyne visibility and the photo-detector efficiency. Measured spectra of the amplitude and phase quadrature variances of the two entangled beams are shown in fig. 6. Both spectra are greater that the quantum noise limit over the entire range of measurement, a necessary prerequisite for entanglement. Due to the symmetric arrangement of our experiment the spectra are identical, so that the assumption of symmetry made in sections III D 1, III D 2, and III D 3 seem reasonable.

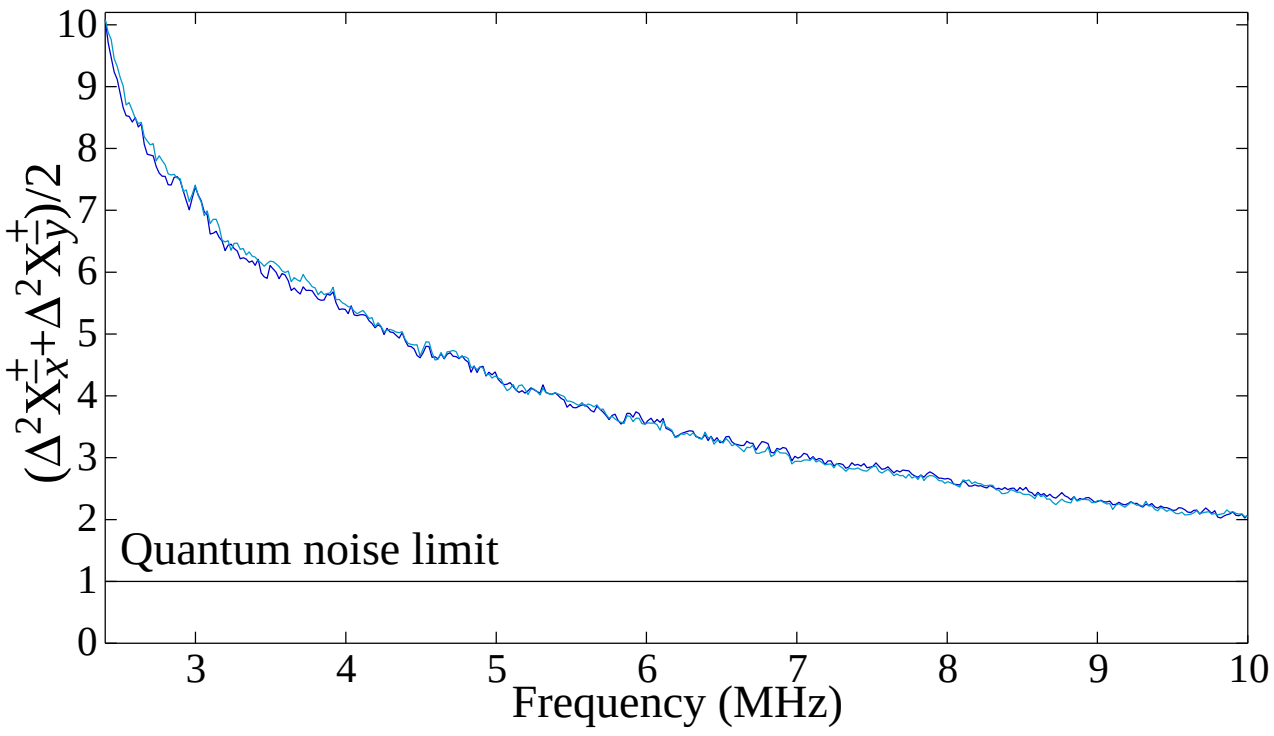

FIG. 6: Frequency spectra of the average amplitude $(\Delta^2\hat{X}^+)$ and phase $(\Delta^2\hat{X}^-)$ quadrature variances of the individual entangled beams normalized to the quantum noise limit.

We analyzed the correlations between beams $x$ and $y$ by measuring the amplitude and phase quadrature sum and difference variances $\Delta^2\hat{X}^{\pm}_{x\pm y}$. The gain between the two homodyne detectors was verified to be unity by encoding large correlated phase modulations on beams $x$ and $y$, throughout the experiment these modulations were suppressed on subtraction by greater than 30 dB. Spectra for $\Delta^2\hat{X}^{\pm}_{x\pm y}$ were then obtained by taking the minimum of the sum and difference variances between homodynes $x$ and $y$ with both homodynes locked to either the amplitude or phase quadratures. These spectra were normalized to the vacuum noise scaled by the combined power of the two homodyne local oscillators and the two entangled beams, and are shown in fig. 7. At frequencies above 5 MHz both the amplitude and phase quadrature sum and difference variances are identical and well below the level expected between a pair of coherent states of the same power. At lower frequencies however, the symmetry between the amplitude and phase quadratures is broken. This effect is due to the relaxation oscillation of the laser which is common mode, and therefore correlated, between the entangled beams. As shown in section II the amplitude quadratures of our entangled beams were anti-correlated, and the phase quadratures were correlated. $\Delta^2\hat{X}^+_{x\pm y}$ was therefore obtained by summing the amplitude quadrature photo-currents from homodynes $x$ and $y$, and the contribution from the relaxation oscillation was therefore also summed. $\Delta^2\hat{X}^-_{x\pm y}$ on the other hand was obtained by subtracting the phase quadrature photo-currents from the homodynes, and so the contributions from the relaxation oscillation cancelled. We see then that with decreasing frequency $\Delta^2\hat{X}^+_{x\pm y}$ degrades quickly, whereas $\Delta^2\hat{X}^-_{x\pm y}$ remains roughly constant. The slight degradation of $\Delta^2\hat{X}^-_{x\pm y}$ at frequencies below 4 MHz can be attributed to small differences in the response of the two homodyne detectors so that the relaxation oscillation was not quite perfectly cancelled.

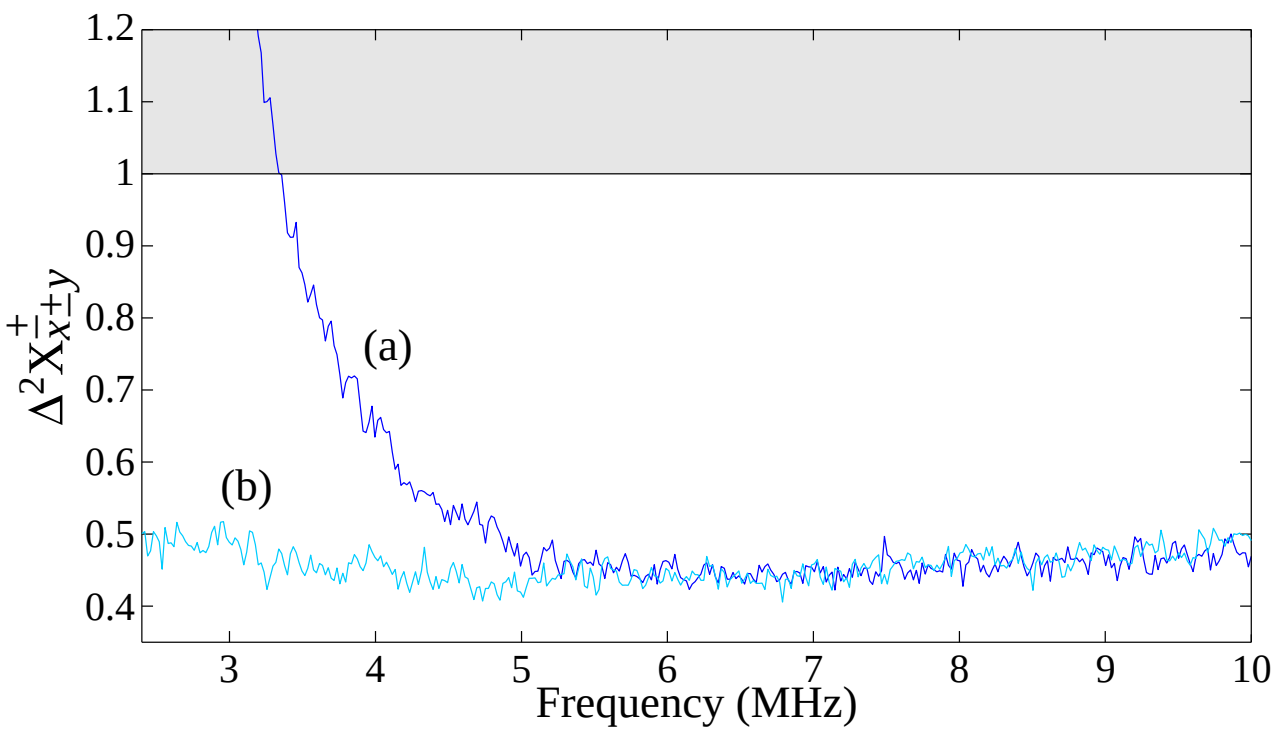

FIG. 7: Frequency spectra of the amplitude and phase quadrature sum and difference variances between beams $x$ and $y$. (a) $\Delta^2\hat{X}^+_{x\pm y}$ and (b) $\Delta^2\hat{X}^-_{x\pm y}$.

### C Characterization of the correlation matrix

As discussed in section III A, the correlation matrix provides a complete characterization of Gaussian entanglement. Given the assumptions that entangled beams $x$ and $y$ are interchangeable and that there are no cross-quadrature correlations the correlation matrix is completely specified through measurements of $C^{\pm\pm}_{xx} = \Delta^2\hat{X}^\pm$ and $C^{\pm\pm}_{xy} = \frac{1}{2}\langle\delta\hat{X}^\pm_x\delta\hat{X}^\pm_y + \delta\hat{X}^\pm_y\delta\hat{X}^\pm_x\rangle$. Measurements of $C^{\pm\pm}_{xx}$ for our entanglement are presented in fig. 6. To obtain $C^{\pm\pm}_{xy}$ we expand $\Delta^2\hat{X}^\pm_{x\pm y}$

$$\Delta^2\hat{X}^\pm_{x\pm y} = \frac{\left\langle\left(\delta\hat{X}^\pm_x \pm \delta\hat{X}^\pm_y\right)^2\right\rangle}{2} \quad (55)$$

$$= \frac{\Delta^2\hat{X}^\pm_x + \Delta^2\hat{X}^\pm_y}{2} \pm \left\langle\delta\hat{X}^\pm_x\delta\hat{X}^\pm_y\right\rangle \quad (56)$$

$$= \Delta^2\hat{X}^\pm \pm \frac{1}{2}\left\langle\delta\hat{X}^\pm_x\delta\hat{X}^\pm_y + \delta\hat{X}^\pm_y\delta\hat{X}^\pm_x\right\rangle \quad (57)$$

$$= C^{\pm\pm}_{xx} \pm C^{\pm\pm}_{xy}. \quad (58)$$

So $C^{\pm\pm}_{xy}$ can be obtained from our measurements of the average amplitude and phase quadrature variances, and the amplitude and phase quadrature sum and difference variances, $C^{\pm\pm}_{xy} = \pm\Delta^2\hat{X}^\pm_{x\pm y} \mp \Delta^2\hat{X}^\pm$. Fig. 8 shows the resulting spectra. We see that $C^{++}_{xy}$ and $C^{--}_{xy}$ are negative and positive, respectively, throughout the range of the measurement. This is a characterization of the correlation and anti-correlation of the phase and amplitude quadratures, respectively, between beams $x$ and $y$.

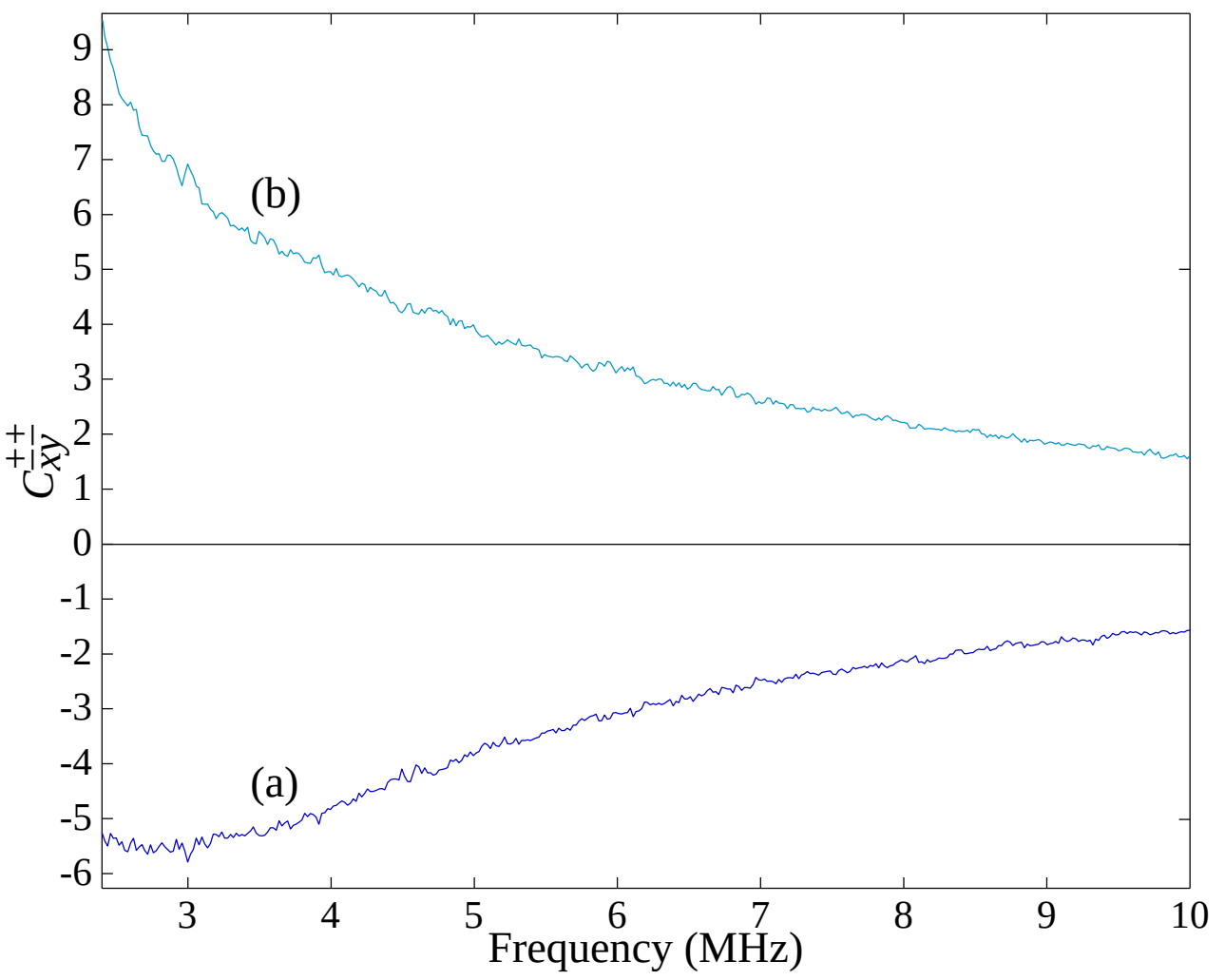

FIG. 8: Frequency spectra of the same-quadrature correlation matrix elements between beams $x$ and $y$. (a) $C^{++}_{xy}$, (b) $C^{--}_{xy}$.

For every sideband frequency, assuming that entangled beams $x$ and $y$ are interchangeable and that there are no cross-quadrature correlations, a correlation matrix describing our entanglement can be constructed from the curves in figs. 6 and 8. Here we take two examples, the correlation matrices of the sidebands at 3.5 and 6.5 MHz. Extracting the data directly from the figures we obtain the correlation matrices

$$CM_{3.5\text{ MHz}} = \begin{pmatrix} 6.2 & 0 & 5.3 & 0 \\ 0 & 6.1 & 0 & 5.7 \\ 5.3 & 0 & 6.2 & 0 \\ 0 & 5.7 & 0 & 6.1 \end{pmatrix}, \quad (59)$$

and

$$CM_{6.5\text{ MHz}} = \begin{pmatrix} 3.3 & 0 & 2.9 & 0 \\ 0 & 3.3 & 0 & 2.9 \\ 2.9 & 0 & 3.3 & 0 \\ 0 & 2.9 & 0 & 3.3 \end{pmatrix}, \quad (60)$$

where all experimentally determined values have an associated statistical error of $\pm 0.05$. The shaded values are fixed as a result of the symmetry assumptions made in section III A and are therefore not experimentally determined. We can now examine whether the inseparability criterion originally proposed by Duan *et al.* (eq. (17)), and the product inseparability criterion of eq. (24) can be used to directly analyse the strength of our entanglement. The correlation matrix given here is of the form required for both criteria (see eq. (14)). It remains, solely, to determine whether the restrictions imposed by each criteria are satisfied. For the original criterion to be valid eqs. (15) and (16) must be true. Since our entangled beams $x$ and $y$ are interchangeable $C^{\pm\pm}_{xx} = C^{\pm\pm}_{yy}$, so that eq. (15) is always true. Eq. (16) on the other hand, is true at 6.5 MHz, but not at the lower frequency of 3.5 MHz. The original inseparability criterion can therefore be used to characterize the strength of our entanglement at 6.5 MHz, but not at 3.5 MHz. For the product criterion to be valid, eq. (24) must be satisfied. Since $C^{\pm\pm}_{xx} = C^{\pm\pm}_{yy}$, for our entanglement at all frequencies, we see that indeed the product criterion is valid for all sideband frequencies. Of course, once the correlation matrix describing the entanglement is fully characterized, it can be transformed into Duan *et al.*'s standard form, and subsequently either inseparability criterion can be used. This, however, involves many more measurements on the entangled state than are required to simply determine the product form of the criterion. Therefore, if a characterization of the inseparability of the entanglement is all that is required, the product form is preferable.

### D Characterization of the inseparability and EPR paradox criteria

A spectrum for the inseparability criterion of eq. (25) was obtained from the amplitude and phase quadrature sum and difference variance spectra in fig. 7. This spectrum is shown in fig. 9. We see that beams $x$ and $y$ were entangled at frequencies within our measurement range higher than 2.8 MHz. As with the other spectra presented in this paper, the strength of the entanglement is degraded at low frequencies as a result of the relaxation oscillation of our laser, and at high frequencies due to the bandwidth of the OPA cavities. The optimum degree of inseparability was achieved at 6.5 MHz, where we observed $\Delta^2\hat{X}^{\pm}_{x\pm y} = 0.44\pm0.01$ for both the amplitude and phase quadratures. This resulted in a degree of inseparability of $\mathcal{I} = 0.44\pm0.01$.

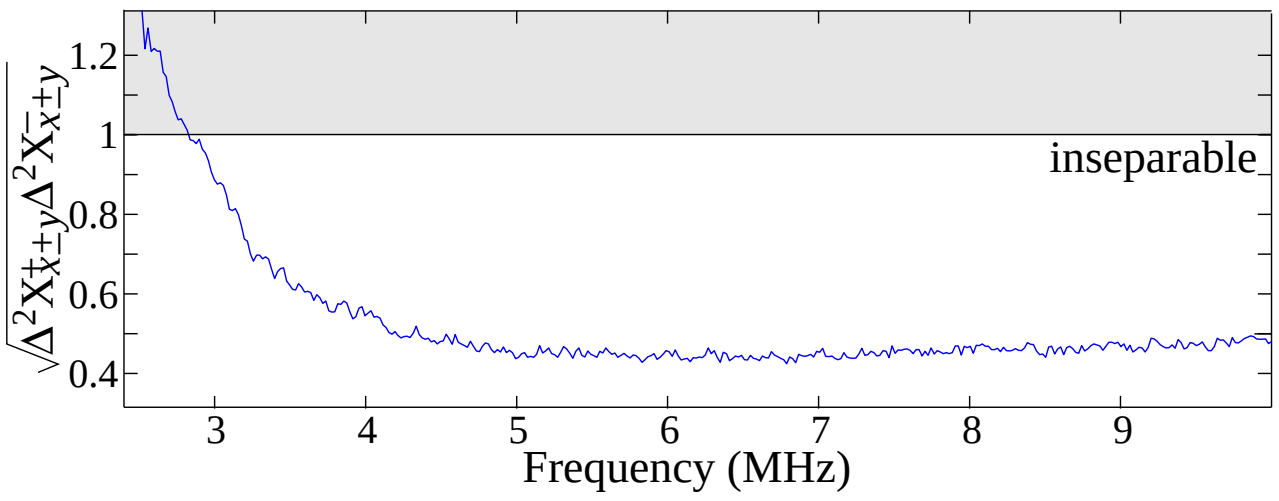

FIG. 9: Frequency spectrum of the degree of inseparability $\mathcal{I}$ between the amplitude and phase quadratures of our entangled state.

Characterization of the EPR paradox criterion requires measurements of the amplitude and phase quadrature conditional variances between beams $x$ and $y$. As can be seen from eq. (29), these variances can be inferred from the correlation matrix elements $C^{\pm\pm}_{xx}$, and $C^{\pm\pm}_{xy}$. However, since these conditional variances were easily measurable from our experimental setup, we measured them directly. The conditional variance measures the uncertainty of one variable ($\hat{X}^{+}_{x}$ say) given knowledge of another variable ($\hat{X}^{+}_{y}$ say). We characterize it here in a similar manner to that used to characterize the sum and difference variances. This time, however, rather than being fixed to unity, the gain between the two homodyne photocurrents was optimized to minimize the measured variances; and the normalization was performed with respect to vacuum fluctuations scaled by only one homodyne local oscillator and entangled beam. The resulting amplitude and phase quadrature conditional variance spectra are shown in fig. 10. We see that both $\Delta^2\hat{X}^{\pm}_{x|y}$ are below unity for the majority of our measurement range. This implies that a measurement performed on beam $y$ will prepare beam $x$ in a squeezed state, and therefore that non-classical correlations exist between the two beams. At 6.5 MHz we obtained the conditional variances $\Delta^2\hat{X}^{+}_{x|y} = 0.77\pm0.01$ and $\Delta^2\hat{X}^{-}_{x|y} = 0.76\pm0.01$. Notice that again, the amplitude quadrature spectrum is strongly degraded at low frequencies due to the relaxation oscillation of our laser, whereas the phase quadrature is unaffected by it.

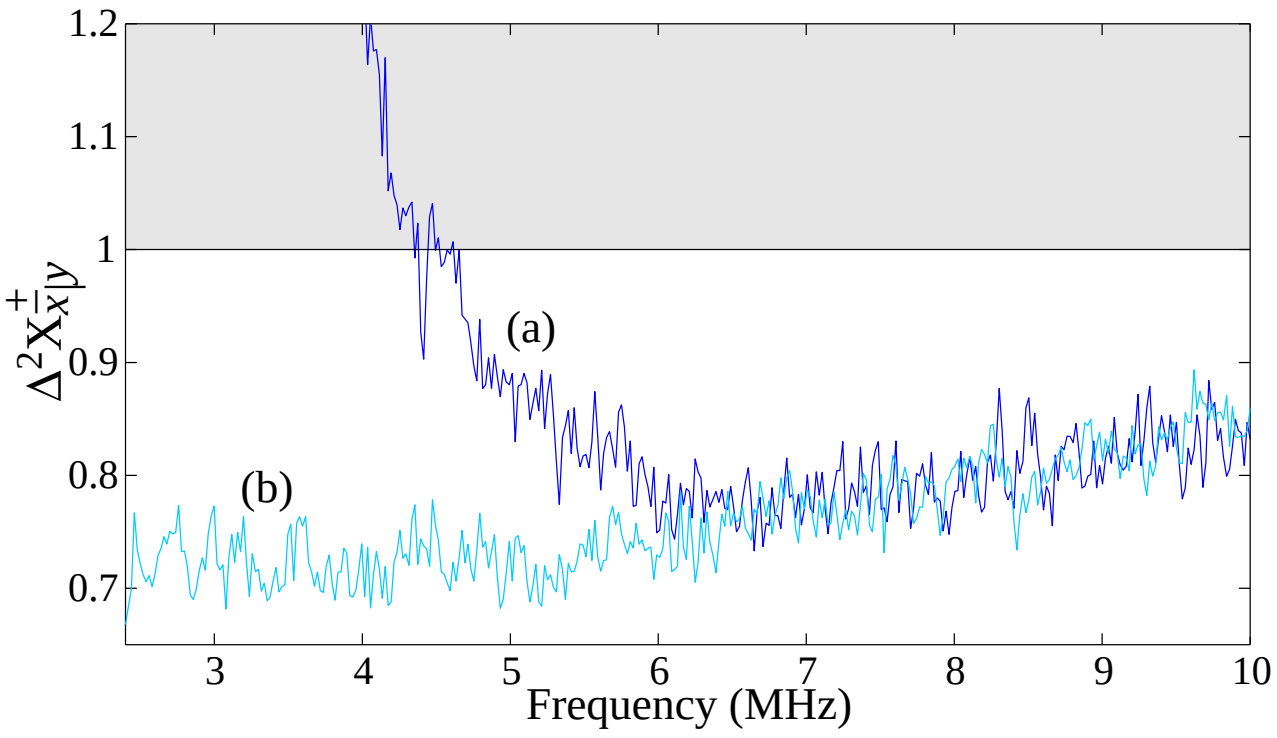

FIG. 10: Conditional variance of the amplitude (a) and phase (b) quadratures of beam $x$ given a measurement on beam $y$ of that quadrature.

Taking the product of the amplitude and phase quadrature conditional variances yields the degree of EPR paradox. Fig. 11 presents the resulting frequency spectrum. We observe an optimum of $\mathcal{E} = 0.58\pm0.02 < 1$, which is well within the regime for observation of the EPR paradox.

We know from the discussion in section III that the degree of EPR paradox $\mathcal{E}$ is highly sensitive to entanglement impurity, whereas the degree of inseparability $\mathcal{I}$ is independent of it. We interrogate this qualitative difference by introducing equal loss to the two entangled beams. Each entangled beam was passed through a waveplate and polarizing beam splitter before detection as shown in fig. 4. Rotating the waveplate allowed us to vary the amount of loss introduced. We characterized both the degree of EPR paradox and the degree of Inseparability at 6.5 MHz for a number of loss settings (waveplate settings). For each measurement the spectrum analyzer was set to zero span and averaged over ten consecutive traces. Fig. 12 summarizes these measurements. We see that the experimental dependences on loss for both $\mathcal{E}$ and $\mathcal{I}$ agree very well with the theoretical curves obtained from eqs. (26) and (31). As discussed in [13], no matter what the loss, the inseparability criterion always holds. We find however, that the EPR paradox criterion fails for loss greater than 0.48. In fact as observed earlier, it is impossible for the EPR paradox criterion to hold for loss greater than or equal to 0.5. The error bars on the plots can be attributed to uncertainty in the loss introduced, small fluctuations in the local oscillator powers and, for the EPR paradox criterion, error in the optimization of the electronic gain.

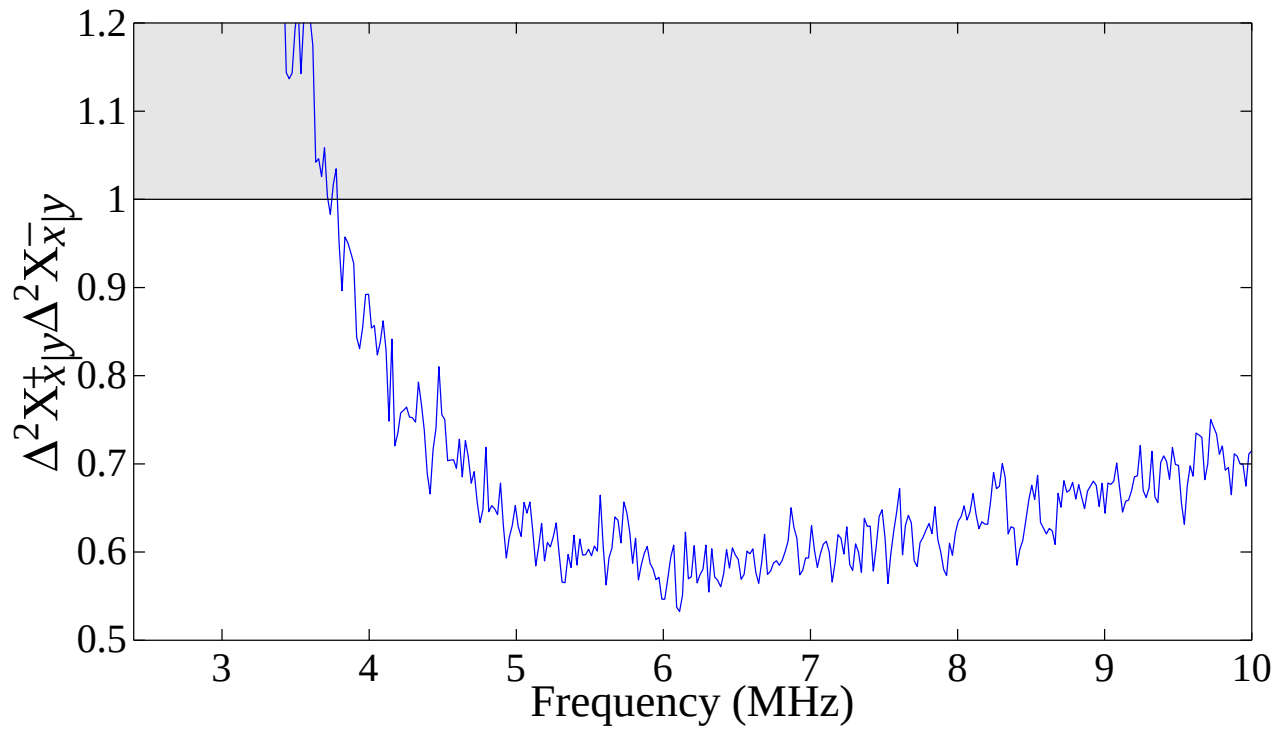

FIG. 11: Frequency spectrum of the degree of EPR paradox between the amplitude and phase quadratures of our entangled state.

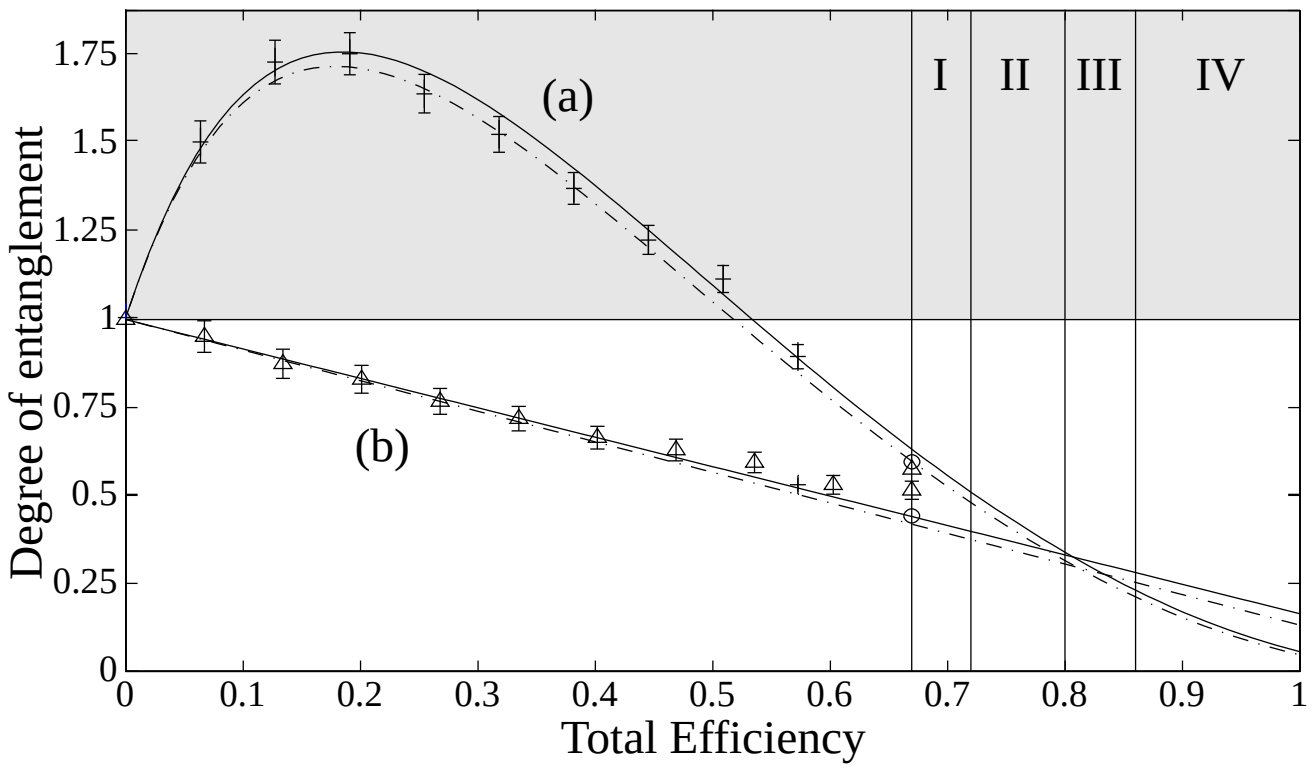

FIG. 12: Comparison of (a) EPR and (b) inseparability criteria with varied detection efficiency. The symbols $+$, $\triangle$, and $\bigcirc$ label three separate experimental runs. For $+$ a systematic error was introduced by the detection darknoise when optimizing the EPR paradox criterion gain. The solid fit in (a) includes this, the dashed fit is the result expected if the error was eliminated, and agrees well with runs $\triangle$ and $\bigcirc$. The solid line in (b) is a theoretical fit, the dashed line is the result predicted by the fit in (a). There were four sources of unavoidable loss in our system, I: Detection loss, II: Homodyne loss, III: optical loss and IV: OPA escape loss.

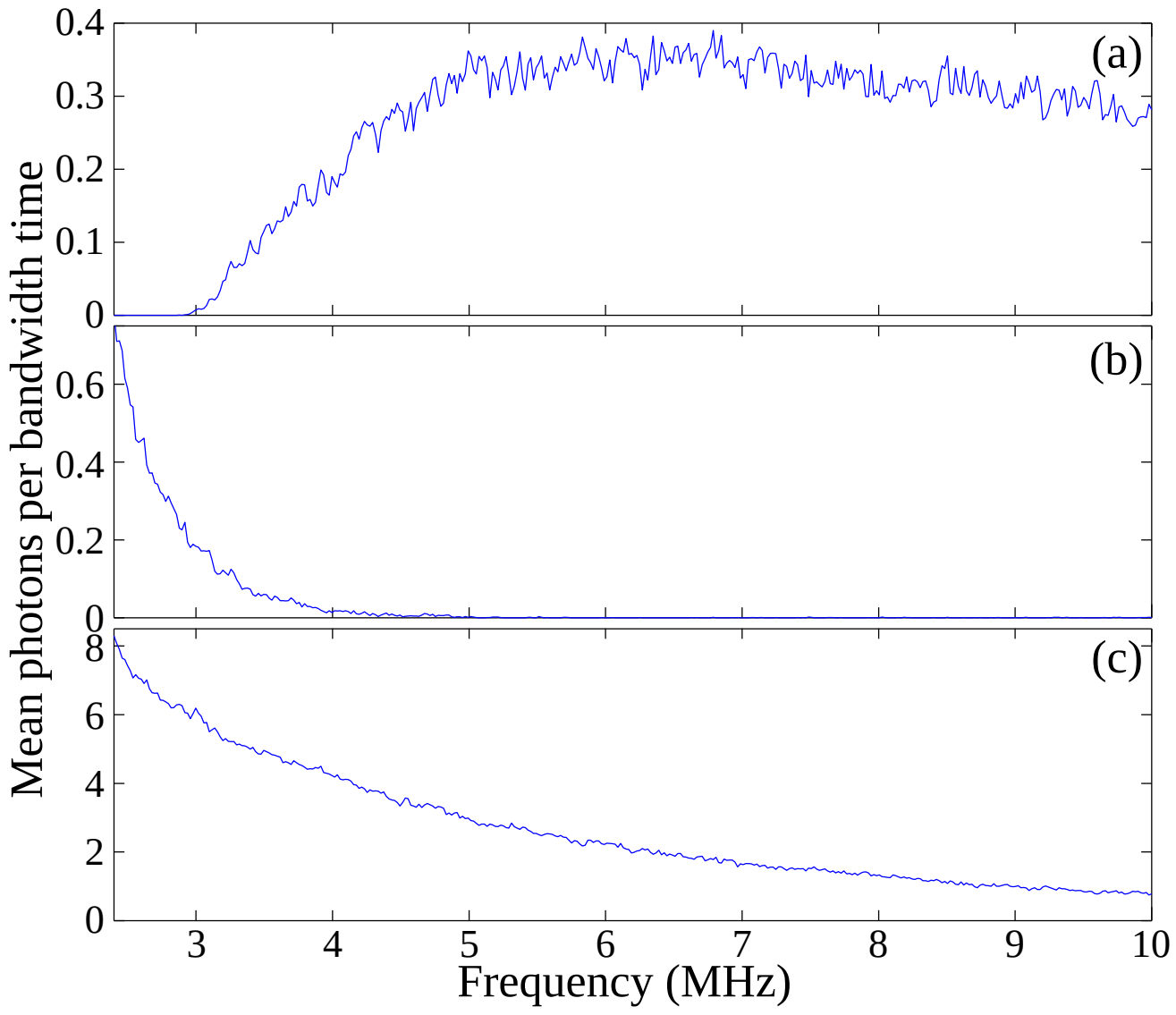

FIG. 13: Frequency spectra of the axes of the photon number plot. (a) $\bar{n}_{\rm min}$, (b) $\bar{n}_{\rm bias}$, and (c) $\bar{n}_{\rm excess}$

### E Representation of results on the photon number diagram

The photon number diagram introduced in section III D and [7] provides a physically intuitive representation of continuous variable entanglement. The measured spectra for $\Delta^2\hat{X}^{\pm}$ and $\Delta^2\hat{X}^{\pm}_{x\pm y}$ shown in figs. 6 and 7 may be translated into the three axes of this diagram ($\bar{n}_{\rm min}$, $\bar{n}_{\rm excess}$, and $\bar{n}_{\rm bias}$) using eqs. (36), (37), (38), and (40). The resulting spectra are shown in fig. 13. At low frequencies there is no entanglement, and from fig. 13 (a) we see that correspondingly no photons are required to maintain the entanglement ($\bar{n}_{\rm min} = 0$), with increasing frequency the average number of photons required increases, peaking at $\bar{n}_{\rm min} = 0.35$ around 6.5 MHz, before dropping off as the frequency moves above the bandwidth of our OPAs. From fig. 13 (b) we see that over the majority of the measured spectrum on average very few photons are present in the entanglement as a result of bias between the amplitude and phase quadratures. Photons resulting from bias only become significant at frequencies below 5 MHz. This bias is a direct consequence of the sensitivity and immunity of $\Delta^2\hat{X}^+_{x\pm y}$ and $\Delta^2\hat{X}^-_{x\pm y}$, respectively, to our lasers relaxation oscillation. Fig. 13 (c) shows that throughout the spectrum of our measurement the majority of the photons present in our entanglement are there as a result of impurity. In fact from the fit to the degree of EPR paradox in fig. 12 we see that at 6.5 MHz the most significant contribution to the impurity of our entangled state is optical loss. Therefore even relatively small levels of loss (such as 33%) facilite a significant transfer of mean photons per bandwidth per time from $\bar{n}_{\rm min}$ to $\bar{n}_{\rm excess}$. If additional sources of phase noise, such as guided-acoustic-wave Brillouin scattering for fibre squeezing [40, 41], are present in the process used to generate squeezing, the average num-ber of photons present due to impurity can become extremely large. The spectra of $\bar{n}_{\rm min}$, $\bar{n}_{\rm excess}$, and $\bar{n}_{\rm bias}$ obtained for our entanglement are mapped onto the photon number diagram in fig. 14.

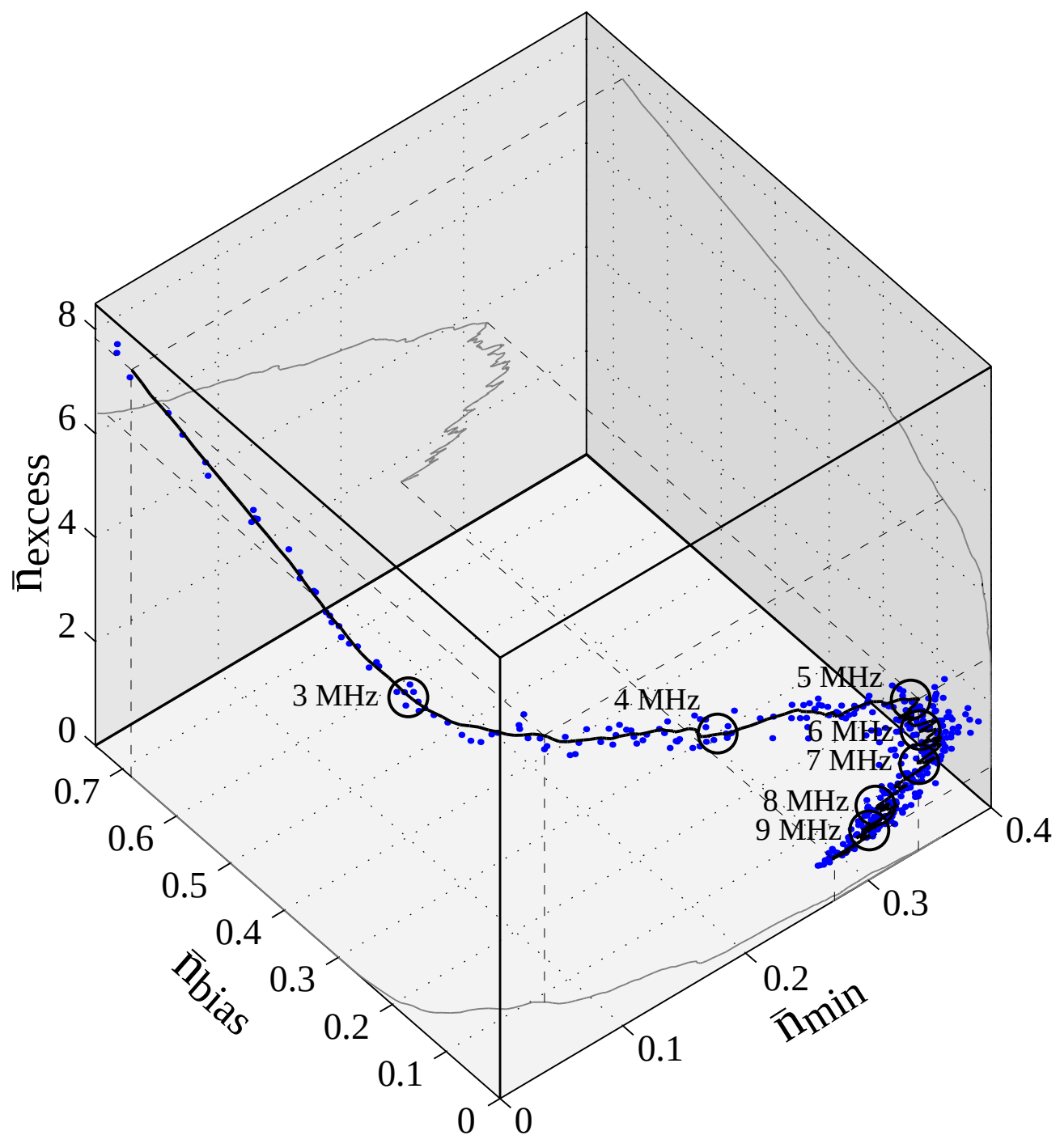

FIG. 14: Representation of the entangled state on the photon number diagram.

The photon number diagram can be used to analyze the efficacy of an entangled state in quantum information protocols. As discussed in section III D, fig. 15 shows efficacy contours of the degree of EPR paradox, quantum teleportation, and high and low photon number dense coding, on the $\bar{n}_{\rm min}$-$\bar{n}_{\rm excess}$ plane of the photon number diagram assuming that $\bar{n}_{\rm bias} = 0$. Since $\bar{n}_{\rm bias} \approx 0$ for our entangled state over most of the measured spectrum, we project the curve shown in fig. 14 onto the $\bar{n}_{\rm bias} = 0$ plane and display it on fig. 15. We can then obtain estimates of the optimum efficacy that could be achieved with our entangled state in various quantum information protocols, and estimates of the frequencies at which the optima occur. From fig. 15 (a) we find that the optimum expected degree of EPR paradox for our entanglement is roughly $\mathcal{E} = 0.68$ and occurs around 6.6 MHz. In section IV D we experimentally obtained a value of $\mathcal{E} = 0.58 \pm 0.02$ which is significantly lower. This difference is evident because the experiment was operating more effectively when the measurements of the degree of EPR paradox were made. Indeed this can be seen in fig. 12, where the degree of inseparability predicted from our the degree of EPR paradox results is somewhat better than the result we obtained directly. Due to sensitivity of the degree of EPR paradox to loss and impurity, this difference completely explains the discrepancy. From fig. 15 (b) we see that the optimum teleportation fidelity achievable with our entanglement is approximately $\mathcal{F} = 0.695$ and would be observed near 6.2 MHz. The entangled state analyzed here was recently used to perform quantum teleportation, due to non-ideal effects such as optical loss and detector darknoise an optimum fidelity of $\mathcal{F} = 0.64 \pm 0.02$ was observed [39]. The low photon number efficacy contours for dense coding shown in fig. 15(c) have an extremely strong dependence on the average number of excess photons carried by the entanglement, accordingly the optimum ratio of dense coding to squeezed state channel capacities would occur at 10 MHz where our entanglement is most pure, in our case this never exceeds unity. However, as discussed in section III D 3, increasing the total average number of photons allowed in the sidebands ($\bar{n}_{\rm encoding}$) causes the dense coding protocol to become independent of $\bar{n}_{\rm excess}$. We find that when a large number of photons per bandwidth per time are available to encode signals ($\bar{n}_{\rm encoding} \gg \bar{n}_{\rm excess}$) the optimum achievable ratio of channel capacities is $C_{\rm EPR}/C_{\rm sqz} \approx 1.02$ and occurs near 6.3 MHz. So that in the large photon number limit dense coding using the entangled state characterized in this paper could yield a channel capacity marginally better than that achievable with optimal squeezed state encoding.

## V. CONCLUSION

In conclusion, we have generated a strongly quadrature entangled state from amplitude squeezed beams produced in two independent OPAs. The correlation matrix of the state was characterized. We gauged the strength of the entanglement in the spirit of the Schrödinger picture by measuring the degree of inseparability, and in the spirit of the Heisenberg picture by measuring the degree of EPR paradox, with optimum results of $\mathcal{I} = 0.44 \pm 0.01$ and $\mathcal{E} = 0.58 \pm 0.02$, respectively. Through the introduction of controlled loss to each entangled beam, qualitative differences between the behavior of the degree of inseparability and the degree of EPR paradox were demonstrated. We characterized the entanglement on a photon number diagram which provides an intuitive and physically meaningful representation of the state. On this diagram the average number of photons per bandwidth per time in the entangled state is separated into components required to maintain the strength of the entanglement, the bias between the amplitude and phase quadratures of the state, and the states impurity. We calculated efficacy contours for the degree of EPR paradox, quantum teleportation and dense coding protocols on the photon number diagram, and used them to predict the level of success achievable for each protocol using our entanglement.

## VI. ACKNOWLEDGEMENTS

This work was supported by the Australian Research Council and is part of the EU QIPC Project, No. IST-1999-13071

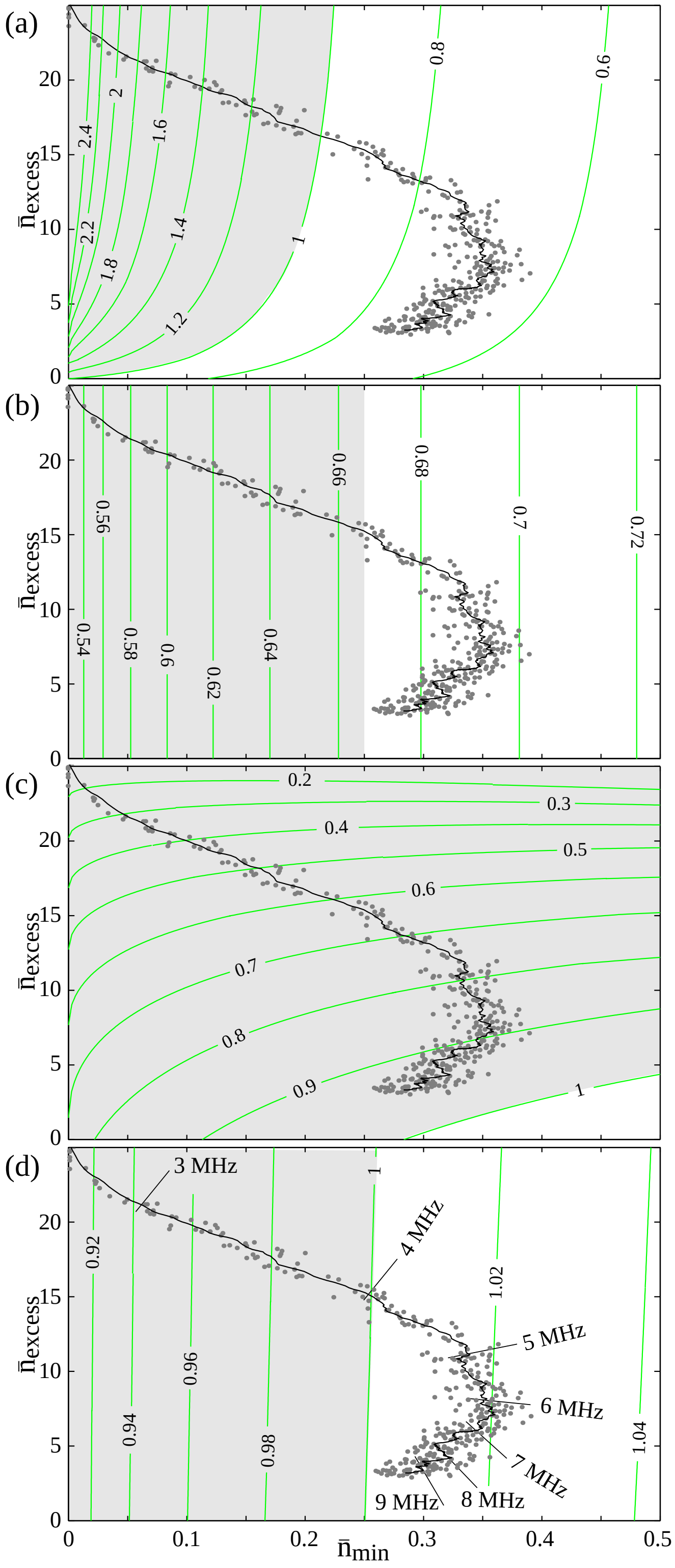

FIG. 15: Two dimensional slice of the photon number diagram for $\bar{n}_{\rm bias} = 0$. The contours on the plots are (a) the degree of EPR paradox, (b) the fidelity of quantum teleportation, (c) and (d) ratio of dense-coding channel capacity to optimum squeezed channel capacity for $\bar{n}_{encoding} = 3.375$ and $\bar{n}_{encoding} = 125$, respectively.

(QUICOV). R. S. acknowledges the Alexander von Humboldt foundation for support. We are grateful of N. Treps, B. C. Buchler, T. Symul and H. A. Bachor for invaluable advice.